\documentclass[11pt]{article}
 \usepackage{graphicx}

\usepackage{graphics,graphicx,psfrag,color,float, hyperref}
\usepackage{epsfig,bbold,bbm}
\usepackage{bm}
\usepackage{mathtools}
\usepackage{amsmath}
\usepackage{amssymb}
\usepackage{amsfonts}
\usepackage{amsthm, hyperref}
\usepackage{amsfonts}
\usepackage {times}
\usepackage{bbm}
\usepackage{soul}
\usepackage{titling}
\usepackage{amsmath}
\usepackage{layout}
\usepackage{amssymb}
\usepackage{amsfonts}
\usepackage{amsthm, hyperref}
\usepackage {times}
\usepackage{setspace}
\usepackage{amsfonts}

\newtheorem{fact}{Fact}

\newcommand{\beq}{\begin{equation}}
\newcommand{\enq}{\end{equation}}
\newcommand{\bel}{\begin{lemma}}
\newcommand{\enl}{\end{lemma}}
\newcommand{\bet}{\begin{theorem}}
\newcommand{\ent}{\end{theorem}}

\newcommand{\tr}{\mathrm{Tr}}

\newcommand{\rv}[1]{\mathbf{#1}}
\newcommand{\hilbertspace}[1]{\mathcal{#1}}
\newcommand{\universe}[1]{\mathcal{#1}}

\newcommand{\E}{\mathbb{E}}

\newcommand{\ketbra}[1]{|#1\rangle\langle#1|}

\newcommand{\ceil}[1]{\left\lceil #1 \right\rceil}

\newcommand{\eps}{\varepsilon}
\newcommand*{\cC}{\mathcal{C}}
\newcommand*{\cA}{\mathcal{A}}

\newcommand*{\cH}{\mathcal{H}}
\newcommand*{\cM}{\mathcal{M}}

\newcommand*{\cN}{\mathcal{N}}
\newcommand*{\cS}{\mathcal{S}}

\newcommand*{\cT}{\mathcal{T}}
\newcommand*{\cX}{\mathcal{X}}

\newcommand*{\cZ}{\mathcal{Z}}
\newcommand*{\cE}{\mathcal{E}}

\newcommand*{\bV}{\mathbf{V}}

\newcommand*{\oI}{\overline{I}}
\newcommand*{\bZ}{\mathbf{Z}}

\newcommand*{\cV}{\mathcal{V}}
\newcommand*{\cY}{\mathcal{Y}}

\newcommand*{\bY}{\mathbf{Y}}

\newcommand{\bra}[1]{\langle #1|}
\newcommand{\ket}[1]{|#1 \rangle}

\mathchardef\mhyphen="2D
\newcommand*{\uI}{\underline{I}}

\newcommand*{\renyi}{R\'{e}nyi }

\makeatletter
\newcommand*{\rom}[1]{\expandafter\@slowromancap\romannumeral #1@}
\makeatother

\mathchardef\mhyphen="2D

\newtheorem{definition}{Definition}
\newtheorem{claim}{Claim}

\newtheorem{theorem}{Theorem}
\newtheorem{lemma}{Lemma}
\newtheorem{corollary}{Corollary}
\newtheorem{proposition}{Proposition}

\begin {document}
\title{One-Shot Private Classical Capacity of Quantum Wiretap Channel:  Based on one-shot quantum covering lemma}
\author{ Jaikumar Radhakrishnan
\thanks { Tata Institute of Fundamental Research, Mumbai,
Email: {\sf jaikumar@tifr.res.in 
}
}
\and 
Pranab Sen
\thanks{
Tata Institute of Fundamental Research, Mumbai,
Email: 
{\sf pranab.sen.73@gmail.com 
}
}
\and 
Naqueeb Ahmad Warsi 
\thanks{ SPMS, NTU, CQT, NUS, Singapore, IIITD, Delhi
Email:
{\sf  naqueeb@iiitd.ac.in
 }
}
}

\date{}
\maketitle
\begin{abstract}
In this work we study the problem of communication over the quantum wiretap channel. For this channel there are three parties Alice (sender), Bob (legitimate receiver) and Eve (eavesdropper). We obtain upper and lower bounds on the amount of information Alice can communicate to Bob such that Eve gets to know as little information as possible about the transmitted messages. Our bounds are in terms of quantum hypothesis testing divergence and smooth max quantum relative entropy. To obtain our result we prove a one-shot version of the quantum covering lemma along with operator Chernoff bound for non-square matrices. 
\end{abstract}

\section{lntroduction}
In this work we consider the problem of communication over a quantum wiretap channel with one sender (Alice) and two receivers (Bob and Eve). They have access to a channel that takes one input $X$ (supplied by Alice) and produces two outputs $Y$ and $Z$, received by Bob and Eve respectively. The characteristic of the channel is given by $p_{YZ \mid X}$ . The goal is to obtain bound on the amount of information Alice may can communicate to Bob such that Eve gets to know as little information as possible about the transmitted messages.

\paragraph{Secrecy capacity of a sequence of wiretap channels in the information spectrum setting \cite{arbitrary-wiretap}:} Bloch and Laneman studied the above problem in the classical asymptotic non-iid setting (information spectrum) wherein they defined various measures of secrecy \cite{arbitrary-wiretap}. One such secrecy measure is the $L_1$ distance $\|p_{MZ}-p_{M}\times p_Z\|$, where the distribution $p_{MZ}$ represents the joint distribution between the transmitted message random variable and the channel output at the Eve's end when Alice transmits $M$. To place our contributions in place, it will be useful to revisit the result of Bloch and Laneman. But we need the following definitions to state Bloch and Laneman's result.

\begin{definition}
An $(n,R,\eps_n,\delta_n)$-wiretap code for a sequence of wiretap channels $\bm{p_{YZ \mid X}}$ consists of the following:
\begin{itemize}
\item a message set $\cM_n:=\left\{1,\cdots, 2^{nR}\right\}$;
\item a stochastic encoding function $e_n : \cM_n \to \cX^n$;
\item a decoding function $d_n : \cY^n \to \cM_n$.
\end{itemize}
The rate of the code is defined as $\frac{1}{n}\log|\cM_n|$. Let $M$ be the random variable denoting the uniform choice of message $i \in \cM_n,$ and $Y^n(Z^n)$ be the random variable representing the legitimate receiver (eavesdropper) output corresponding to $e_n(M)$. The average probability of error is defined as $\eps_n = \Pr\left\{d_n(Y^n) \neq M\right\}$ and the secrecy is measured in terms of $\left\|p_{MZ}-p_{M}\times p_Z\right\|.$
\end{definition}
\begin{definition}
A rate $R$ is achievable for a sequence of wiretap channels $\bm{p_{YZ \mid X}}$ if there exists a sequence of $(n, R, \eps_n, \delta_n)$-wiretap code such that $\lim_{n \to \infty}\eps_n = 0$ and $\lim_{n \to \infty}\delta_n = 0$.
\end{definition}
The supremum of all such achievable rates is called the private capacity of the sequence wiretap channels $\bm{p_{YZ \mid X}}$ and we represent it by $P$.

\begin{definition}(Specrtal inf-mutual information rate \cite{han-book})
Let $\mathbf{V}= \left\{V^n\right\}_{n=1}^{\infty}$ and $\mathbf{Y}= \left\{Y^n\right\}_{n=1}^{\infty}$ be two sequences of random variables where for every $n$, $V^n \in \cV^n$, $Y^n \in \cY^n$ and $(V^n,Y^n) \sim p_{V^nY^n}$. The spectral inf-mutual information $\uI(\bV;\bY)$ is defined as follows
\beq
\underline{I}(\bV;\bY):=\sup\left\{\beta: \lim_{n\to \infty} \Pr\left\{\frac{1}{n}\log\frac{p_{V^nY^n}}{p_{V^n}p_{Y^n}} < \beta\right\} = 0\right\}. \nonumber
\enq
The probability on  the R.H.S. of the above equation is calculated with respect to the distribution $p_{V^nY^n}$.
\end{definition}
\begin{definition}(Specrtal sup-mutual information rate \cite{han-book})
Let $\mathbf{V}= \left\{V^n\right\}_{n=1}^{\infty}$ and $\mathbf{Z}= \left\{Z^n\right\}_{n=1}^{\infty}$ be two sequences of random variables where for every $n$, $V^n \in \cV^n$, $Z^n \in \cZ^n$ and $(V^n,Z^n) \sim p_{V^nZ^n}$. The spectral sup-mutual information $\uI(\bV;\bZ)$ is defined as follows
\beq
\oI(\mathbf{V};\mathbf{Z}):=\inf\left\{\beta: \lim_{n\to \infty} \Pr\left\{\frac{1}{n}\log\frac{p_{V^nZ^n}}{p_{V^n}p_{Z^n}} > \beta\right\} = 0\right\}. \nonumber
\enq
The probability on  the R.H.S. of the above equation is calculated with respect to the distribution $p_{V^nZ^n}$.
\end{definition}

We now state the result of Bloch and Laneman  \cite{arbitrary-wiretap}.
\begin{theorem} \cite{arbitrary-wiretap}
\label{blochandlaneman}
Let $\bm{p_{YZ \mid X}}:=\left\{p_{Y^nZ^n\mid X^n}\right\}_{n=1}^{\infty}$ represent a sequence of wiretap channels. The secrecy capacity $\left(P\right)$ for this sequence of channels is the following:
\beq
P = \max_{(\bm{V},\bm{X})} \left(\underline{I}\left[\bm{V};\bm{Y}\right]-\overline{I}\left[\bm{V};\bm{Z}\right]\right),
\enq
where $(\bm{V},\bm{X})$ represents a sequence of pair of random variables $\left\{V^n,Z^n\right\}_{n=1}^\infty.$
\end{theorem}

A quantum version of the wiretap channel was studied by Devetak in \cite{devetak-2005} and 
Cai-Winter-Yeung in \cite{cai-winter-yeung}, where instead of $p_{YZ \mid X}$, the channel is characterised by the map $\cN^{\cA \to BE} : \cS(\cH_A) \to \cS(\cH_{BE})$. They showed the following.
\begin{theorem}
The private classical capacity of a quantum channel $\cN^{A \to BE}$ in the asymptotic iid setting is the following:
\beq
\lim_{k \to \infty}\frac{1}{k} P(\cN^{\otimes k}), \nonumber
\enq
where $P(\cN)$ is defined as 
\beq
P(\cN) := \max_{\rho}\left(I[V;B]_{\sigma}-I[V;E]_{\sigma}\right), \nonumber
\enq
where all the information theoretic quantities are calculated with respect to the following state:
$$\sigma^{VBE} = \sum_{v \in \cV}p_{V}(v)\ket{v}\bra{v}^V\otimes \cN^{A \to BE}(\rho^A_v).$$
\end{theorem}

\section{Our Result}
\begin{figure}[H]
\centering
\resizebox{0.8\textwidth}{!}{
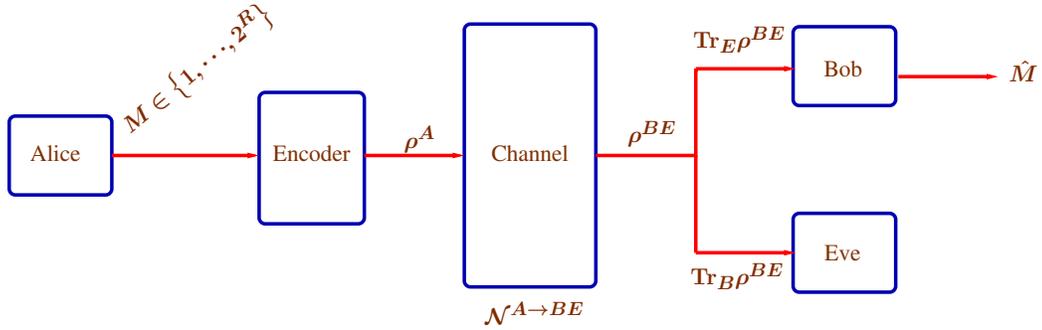
}
\vspace{0.1cm}
\caption{Private classical information transmission model over 
a quantum channel}
\label{wire}
\end{figure}

We consider the above problem in the quantum one-shot setting. A quantum wiretap channel takes a quantum input $\rho^A$ and produces two quantum outputs $\rho^B$ and $\rho^E$, received by Bob and Eve respectively. The characteristics of the channel is given by $\cN^{A \to BE}(\rho^A) = \rho^{BE}$. A communication scheme 
over a quantum wiretap channel is illustrated in Figure \ref{wire} .
We need the following definition to discuss our results.
\begin{definition}[Encoding, Decoding, Error, Secrecy]
\label{wiretapcode}
A one-shot $(R, \eps, \delta)$-code for a quantum wiretap channel consists of 
\begin{itemize}
\item an encoding function $F: [2^{R}] \rightarrow \mathcal{S}(\mathcal{H}_A)$, such that 
\beq
\left\|\rho^{ME} -{\rho}^{M}\otimes{\rho}^E\right\| \leq \delta,
\enq
where $\rho^{ME}$, $\rho^M$ and $\rho^E$ are  
appropriate marginals of the state $\rho^{MBE}=\frac{1}{2^{R}}\sum_{m \in [2^R]}\ket{m}\bra{m}\otimes\cN^{A \to BE}(F(m))$.
\item decoding POVMs $\{\cT^B_{m}: m \in [2^{R}]\}$ such that the average probability
of error
\beq
\frac{1}{2^{R}}\sum_{m\in [2^{R}]}
p_e(m) \leq \eps,
\enq
where 
\beq
p_e(m) =\tr\left[\left(\mathbb{I} -\cT^B_{m}\right)\cN(F(m))\right]
\enq
is the probability of error when Alice uses this scheme to transmit the message $m$.
\end{itemize}
\end{definition}
\begin{definition}(Quantum hypothesis testing divergence \cite{wang-renner-prl})
\label{mindiv}
Let $\rho^{VB}:=\sum_{v\in\cV}p_{V}(v)\ket{v}\bra{v}^U\otimes\rho_v^B$ be a classical quantum state. For $\eps\in[0,1)$ the hypothesis testing divergence between the systems $V$ and $B$ is defined as follows:
\beq
I^{\eps}_{0}[V;B]:=\sup_{\substack{0\preceq\Gamma\preceq\mathbb{I}\\ \tr\left[\Gamma\rho^{VB}\right]\geq 1-\eps}}-\log\tr\left[\Gamma\left(\rho^{V}\otimes\rho^B\right)\right]. \nonumber
\enq
\end{definition}
\begin{definition}(Quantum smooth max \renyi divergence)
Let $\rho^{VE}:=\sum_{v\in\cV}p_{V}(v)\ket{v}\bra{v}^V\otimes\rho_v^E$ be a classical quantum state. For $\eps\in[0,1)$ the smooth max \renyi divergence between the systems $V$ and $E$ is defined as follows:
\beq
I^{\eps}_{\infty}[V;E]:=\inf\left\{\gamma: \sum_{v \in \cV}p_{V}(v)\tr\left[\left\{\rho^E_{v}\succ 2^{\gamma}\rho^E\right\} \rho^E_v\right] \leq \eps \right\}\nonumber,
\enq
where $\rho^E = \tr_{V}\left[\rho^{VE}\right]$ and $\left\{\rho^E_{v}\succ 2^{\gamma}\rho^E\right\}$ is the projector onto the positive Eigen space of the operator $\rho^E_{v} - 2^{\gamma}\rho^E$.
\end{definition}
\begin{theorem}(Achievability)
\label{achievability}
Let $\cN^{A\to BE}$ be a quantum wiretap
channel. Let $V$ be a random variable taking values in
$\cV$ and $F :\cV \to 
\mathcal{S}(\mathcal{H}_A)$. Consider the state 
\beq
\label{jointstate11}
\rho^{VBE} = \sum_{v \in \cV}p_V(v) \ket{v}\bra{v}^V \otimes \cN^{A \to BE}\left(\rho^A_v\right).
\enq
For every $\eps \in (0,1)$ and $\delta \in (0,2)$ there exists an $(R, \eps, \delta)$-code for the quantum wiretap channel $\cN^{A\to BE}$ if 
\begin{align}
\label{achievablerate}
R & \leq  \sup_{\left\{V,F\right\}}\left(I^{\eps'}_{0}[V;B] -\max\left\{0,I^{\hat{\delta}}_{\infty}[V;E]\right\} \right)+ \log\left(\eps'\right)+ \log \left(\hat{\delta}^9\right)-\mathcal{O}\left( \log \log (\dim(\cH_E))\right)
\end{align}
where $18\eps^\prime \leq \eps$ and $\hat{\delta}$ is such that $144\sqrt{\hat{\delta}} \leq \delta$. The information theoretic quantities mentioned in \eqref{achievablerate} are calculated with respect to the state given in \eqref{jointstate11}. 
\end{theorem}


\begin{theorem}(Converse)
\label{converseweak}
For a quantum wiretap channel $\cN^{A \to BE}$ any $(R,\eps, \delta)$-code satisfies the following:
\beq
\label{R}
R \leq \sup_{\left\{V,F\right\}}\left(I^{\eps}_{0}\left[V;B\right] -I^{\delta}_{\infty}\left[V;E\right]\right)+1.5, 
\enq
where $V$ is a random variable over a set $\cV$, 
$F : \cV \to \cS(\cH_A)$ a map from $\cV$ to 
$\mathcal{S}(\mathcal{H}_A)$ and
all the information theoretic quantities are calculated with 
respect to the following state:
\beq
\Theta^{VBE}:= \sum_{v \in \cV}p(v)\ket{v}\bra{v}^{V}
\otimes \cN^{A\to BE}\left(\rho_{v}^{A}\right).\nonumber 
\enq
\end{theorem}
\paragraph{Techniques:} Our achievability proof follows along the 
line of the proof in \cite{wilde-book}. As before, we generate a row array, whose entries are
generated according to the distribution $p_V$; furthermore, as in the original proof we
partition this array into bands of appropriate sizes and uniquely assign each of these bands to a message.  
To send a message $m \in [2^R]$, Alice chooses a codeword $v$ 
uniformly from the band corresponding to $m$; applies the 
map $F$ to $v$ and then transmits the resulting state $\rho^A_v$ over 
the channel. Bob on receiving his share of the channel output tries 
to determine the codeword $v$ using standard one-shot decoding techniques 
for a point to point quantum channel. He succeeds with high probability
for the given codebook size.  It only remains to show that the message
$m$ is secret from Eve. The random choice of $v$ from the band corresponding
to $m$ should make Eve's share of the channel output independent of $m$.
This is the main technical hurdle that must be overcome in order to
prove the correctness of a code for a wiretap channel.
In the asymptotic iid setting, this hurdle is overcome by
proving a {\em quantum covering lemma} \cite[Lemma 16.2.1]{wilde-book} based on an operator Chernoff
bound of Ahlswede-Winter \cite{ahlswede-02-converse} for Hermitian 
matrices . Unfortunately, a straightforward translation of this technique to one-shot
setting fails. In this work, we overcome these difficulties and manage to
prove for the first time a {\em one-shot quantum covering lemma}. 
On the way, we also prove a novel operator Chernoff bound for non-square
matrices.

The proof for the converse (Theorem \ref{converseweak}) essentially follows along the 
line of the proof given in \cite{arbitrary-wiretap}; the translation
to the one-shot quantum setting is straightforward.

\paragraph{Private classical capacity of the quantum wiretap channel in the quantum information spectrum setting.} Our bounds allow us to obtain the quantum version of 
Theorem \ref{blochandlaneman}. The quantum information spectrum technique pioneered by Hayashi
and Nagaoka~\cite{Hayashi-noniid} allows one to derive meaningful
bounds on rates even in the absence of the iid assumption; however,
the analysis is often more challenging in this setting. The bounds
in our work are expressed using smooth min and max \renyi divergences. The close
relationship between these quantities and the quantities that typically
arise in the information spectrum setting (see Datta and Leditzky \cite{datta-2014-secondorder}) allows us to derive the quantum version of Theorem \ref{blochandlaneman}.

\paragraph {Related work:}  In \cite{renes-renner-2011} Renes and 
Renner derive one-shot achievability and converse bounds for the 
quantum wiretap channel in terms of conditional min and max 
\renyi entropies. They also show that their result asymptotically yields 
the results of  \cite{devetak-2005} and \cite{cai-winter-yeung}. 
However, the result of Renes and Renner \cite{renes-renner-2011} 
does not seem to yield the asymptotic characterisation of 
the wiretap channel 
in the information spectrum (non-iid) setting. Such a result as mentioned earlier in Theorem \ref{blochandlaneman} is known for the classical case. We note here that 
our one-shot bounds which are stated in terms of two fundamental smooth \renyi divergences allow us to characterise 
the capacity of the wiretap channel in the information 
spectrum (asymptotic non-iid) setting; our characterisation turns out
to be nothing but the quantum analogue of the characterisation of
Theorem \ref{blochandlaneman}.
\section{Proof of Theorem \ref{achievability}}
\begin{proof}
Let $\rho^{VB}:=\sum_{v\in \cV}p_{V}(v)\ket{v}\bra{v}^{V}\otimes\rho^B_{v}$ denote the joint state of the system $VB$, where $\rho^B_{v}:=\tr_{E}\left[\rho^{BE}_{v}\right]$. Also, let $\rho^B:=\mathbb{E}_{V}\left[\rho_{V}^{B}\right]$. Let $0\preceq \Gamma^{VB}\preceq \mathbb{I}$ satisfy the following properties
\begin{itemize}
\item[(P1)] $\tr\left[\Gamma^{VB}\rho^{VB}\right] >  1- \eps^\prime$.
\item[(P2)] $I^{\eps^\prime}_{0}[V;B] = -\log\tr\left[\Gamma^{VB}\left(\rho^V\otimes\rho^B\right)\right].$
\end{itemize}
See Definition \ref{mindiv} for the definition of $I^{\eps^\prime}_{0}[V;B]$. Fix $R$ such that
$$R \leq I^{\eps'}_{0}[V;B] -\max\left\{0,I^{\hat{\delta}}_{\infty}[V;E]\right\}  + \log\left(\eps'\right)-\log\left(10^{16}\frac{\left(\log\dim(\cH_E)\right)^6}{ \hat{\delta}^9}{\left(-\ln\left(\frac{\hat{\delta}}{30C}\right)\right)}\right),$$
where $C =\dim(\cH_E)\left(\log_2\left( \frac{4\dim(\cH_E)}{\hat{\delta}}\right)+1\right)^2$. Furthermore, let $\tilde{R}$ be such that 
\beq
\label{hatR}
\tilde{R} = \max\left\{0,I^{\hat{\delta}}_{\infty}[V;E]\right\} +\log\left(10^{16}\frac{\left(\log\dim(\cH_E)\right)^6}{ \hat{\delta}^9}{\left(-\ln\left(\frac{\hat{\delta}}{30C}\right)\right)}\right).
\enq
See Definition \ref{qmaxdiv} for the definition of ${I^{\hat{\delta}}_{\infty}[V;E]}.$
\paragraph{Random code generation:}
For every message $m \in [2^R]$, let $\mathcal{C}(m):=\left\{v(m,1),\cdots, v(2^R,k)\right\}$,
denote a collection of $k$ codewords where $k = 2^{\tilde{R}}$, and for every $(m,i) \in [2^R] \times [2^{\tilde{R}}]$, $v(m,i)$ is generated using the distribution $p_{V}$. These $2^{R+\tilde{R}}$ codewords form the codebook denoted by $\cC$. 

\paragraph{Encoding:} To send a message $m \in [2^R]$, Alice chooses a codeword $v(m,i)$ from $\mathcal{C}(m)$ uniformly at random. Alice then applies the function $F$ to $v(m,i)$ and transmits the resulting quantum state $\rho^A_{v}(m,i)$ over the channel.

\paragraph{Decoding:} The decoding strategy we mention here is similar to that mentioned in \cite{wang-renner-prl}. Let us define the following POVM by its element 
\beq
E{(m,i)} := \left(\sum_{(m^\prime,i^\prime) \in [2^R] \times [2^{\tilde{R}}]}\Lambda_{v(m^\prime,i^\prime)}\right)^{-\frac{1}{2}}\Lambda_{v(m,i)}\left(\sum_{(m^\prime,i^\prime) \in [2^R] \times [2^{\tilde{R}}]}\Lambda_{v(m^\prime,i^\prime)}\right)^{-\frac{1}{2}}, \nonumber
\enq
where $\Lambda_{v}:= \tr_{V}\left[\left(\ket{v}\bra{v}^V\otimes \mathbb{I}^B\right)\Gamma^{VB}\right]$. For every $(m,i) \in [2^R] \times [2^{\tilde{R}}]$, Bob is equipped with the decoding POVM $E(m,i)$. Bob on receiving his share of the channel output  uses these decoding POVMs to measure the received quantum state to guess the transmitted message. If the outcome of this measurement is $(m,i)$, Bob declares that the transmitted message was $m$.

%
%
%
\paragraph{Analysis for the probability of error:} We now calculate the average probability of error for the above mentioned  encoding and decoding strategy averaged over all the codebooks. Let $v(M,L)$ denote the codeword chosen for transmitting the message $M$, where $M$ is uniformly distributed over the set $[2^R]$ and $L$ is uniformly distributed over the set $[2^{\tilde{R}}]$.
By the symmetry of the random code construction, the average probability of error is equal to the error probability given the transmission of any specific codeword. Without loss of generality, we assume that the codeword $v(1,1)$ was sent; therefore, letting $\hat{M}$ as the decoded message random variable, we have
\begin{align}
\mathbb{E}_{V}\left[\Pr\{\hat{M}\neq 1\}\right] &\leq \mathbb{E}_{V}\tr\left[\left(\mathbb{I}-E(1,1)\right)\rho_{V(1,1)}\right] \nonumber\\
&\overset{a} \leq 2 \mathbb{E}_V\tr\left[(\mathbb{I}-\Lambda_{V(1,1)})\rho_{V(1,1)}\right]+4 \sum_{(m,i)\neq (1,1)}\mathbb{E}_V\tr\left[\Lambda_{V(m,i)}\rho_{V(1,1)}\right] \nonumber\\
&\overset{b}= 2 \left(1-\tr\left[\Gamma^{VB}\rho^{VB}\right]\right)+ 4 \sum_{(m,i)\neq (1,1)}\tr\left[\Gamma^{VB}\left(\rho^V\otimes\rho^B\right)\right]\nonumber\\
&\overset{c} \leq 2\eps^\prime+ 2^{2+R+\tilde{R}-I^{\eps^\prime}_{0}[V;B]}\nonumber\\
&\overset{d} \leq 6\eps^\prime
\end{align}
where $a$ follows from Hayashi Nagaoka operator inequality; $b$ follows from the definition of $\Lambda_V$; $c$ follows from the definition of $\Gamma^{VB}$ and $d$ follows from our choice of $R$ and $\tilde{R}$. 
\paragraph{Analysis for the leaked information to Eve:} Let $\rho^{ME}:= \sum_{m\in [2^R]}\frac{1}{2^{R}}\ket{m}\bra{m}\otimes\rho^E_{m}$ be the joint state of the system $ME$. Notice that in the setting of the problem for every $m\in [2^R]$
\beq
\label{def11}
\rho^E_{m}=\frac{1}{2^{\tilde{R}}}\sum_{l \in[1:2^{\tilde{R}}]}\rho^{E}_{v(m,l)},
\enq
where $\rho^E_{v}:=\tr_{B}\left[\rho^{BE}_{v}\right]$ and $\rho^{BE}_{v}:=\cN^{A\to BE}(\rho^A_v)$. Furthermore, let $\tilde{\rho}^E=\frac{1}{2^R}\sum_{m =1}^{2^R}\rho^E_{m}$ and ${\rho}^E:=\mathbb{E}_{V}\left[\rho_{V}^{E}\right]$. The leakage information is now calculated as follows: 
\begin{align}
\left\|\sum_{m=1}^{2^R}\frac{1}{2^{R}}\left(\ket{m}\bra{m}\otimes\rho^E_{m}\right)-\sum_{m =1}^{2^R}\frac{1}{2^{R}}\ket{m}\bra{m}\otimes\tilde{\rho}^E\right\|& \leq \sum_{m = 1} ^{2^R}\frac{1}{2^{R}}\left\|\rho^{E}_m -\tilde{\rho}^E\right\| , \nonumber\\
&\leq \sum_{m = 1}^{2^R}\frac{1}{2^{R}}\left\|\rho^{E}_m -{\rho}^E\right\| + \left\|{\rho}^E-\tilde{\rho}^E\right\| \nonumber\\
&\leq 2 \sum_{m=1}^{2^R}\frac{1}{2^{R}}\left\|\rho^{E}_m -{\rho}^E\right\|, \nonumber
\end{align}
where all the above inequalities follow from the triangle inequality.  Therefore, by the symmetry of the random code construction, we have
\begin{align*}
\mathbb{E}_\cC\left\|\sum_{m=1}^{2^R}\frac{1}{2^{R}}\left(\ket{m}\bra{m}\otimes\rho^E_{m}\right)-\sum_{m =1}^{2^R}\frac{1}{2^{R}}\ket{m}\bra{m}\otimes\tilde{\rho}^E\right\| 
&\leq 2 \sum_{m=1}^{2^R}\frac{1}{2^{R}}\mathbb{E}_\cC\left\|\rho^{E}_m -{\rho}^E\right\|, \nonumber\\
&=2\mathbb{E}_{\cC(1)}\left\|\frac{1}{2^{\tilde{R}}}\sum_{l \in[1:2^{\tilde{R}}]}\rho^{E}_{V(1,l)}-{\rho}^E\right\|.
\end{align*}
Thus, we can now conclude from Theorem \ref{Cl} (see Section \ref {CL}) with $\eps \leftarrow \hat{\delta}$ and $I \leftarrow \max\left\{0,I^{\hat{\delta}}_\infty\left[V;E\right]\right\}$, that for our choice of $\tilde{R}$ (see \eqref{hatR}) we have 
\begin{align}
\mathbb{E}_\cC\left\|\sum_{m=1}^{2^R}\frac{1}{2^{R}}\left(\ket{m}\bra{m}\otimes\rho^E_{m}\right)-\sum_{m =1}^{2^R}\frac{1}{2^{R}}\ket{m}\bra{m}\otimes\tilde{\rho}^E\right\| 
& \leq 48\sqrt{\hat{\delta}}. \nonumber
\end{align}

\paragraph{Expurgation:} Let $\eps(\cC)$ be a random variable representing the average probability of error where the randomization is over the random choice of the codebook $\cC$ and let $\delta(\cC)$ be a random variable representing the leakage information and again the randomization is over the random choice of the codebook $\cC$. Further , we define the following events:
\begin{align*}
\cE_1&:=\left\{\eps(\cC) <3\mathbb{E}_{\cC}\Pr\left\{M \neq \hat{M} \right\}\right\}\\
\cE_2 &:=\left\{\delta(\cC) <3\mathbb{E}_\cC\left\|\sum_{m=1}^{2^R}\frac{1}{2^{R}}\left(\ket{m}\bra{m}\otimes\rho^E_{m}\right)-\sum_{m =1}^{2^R}\frac{1}{2^{R}}\ket{m}\bra{m}\otimes\tilde{\rho}^E\right\| \right\}.
\end{align*}
Using Markov inequality it is easy to see from the definition of $\cE_1$ and $\cE_2$ that
\beq
\Pr\left\{\cE_1, \cE_2\right\} > \frac{1}{3}. \nonumber
\enq
Thus, we can now conclude that if $R$ satisfies the condition of the theorem then there exists an $(R,\eps, \delta)$-code for the quantum wiretap channel.
\end{proof}

\section{Proof of Theorem \ref{converseweak}}
We need the following key lemma which can be considered as the quantum generalization of \cite[eq. 2.3.18, p.18]{Pinsker-book}.
\begin{lemma}
\label{pinsker}
Let $\rho$ and $\sigma$ be two quantum states. Furthermore, let $\Pi := \{\rho \succ 2^{\beta}\sigma\}$
where $\beta >0$ is arbitrary. Then,
\beq
||\rho-\sigma|| \geq \frac{2\beta\ln2}{\beta\ln2+1}\tr[\Pi \rho].
\enq
\end{lemma}
\begin{proof}
The proof follows the fact that $\|\rho-\sigma\|\geq 2 \tr\left[\Pi(\rho-\sigma)\right]$, \cite[Lemma 9.1.1]{wilde-book}. The claim now follows from the following set of inequlities,
\begin{align*}
\|\rho-\sigma\| &\geq 2 \tr\left[\Pi(\rho-\sigma)\right]\\
& \overset{a}\geq 2\tr\left[\Pi\rho\right]\left(1- e^{-\beta\ln2}\right)\\
& \overset{b}\geq 2\tr\left[\Pi\rho\right]\left(1 - \frac{1}{1+\beta\ln2}\right)\\
&= \frac{2\beta\ln2}{\beta\ln2+1}\tr\left[\Pi\rho\right],
\end{align*}
where $a$ follows because $\Pi\rho\Pi\succ2^\beta\Pi\sigma\Pi$ and $b$ follows because $e^{\beta\ln2} \geq (1+\beta\ln2)$. This completes the proof.
\end{proof}

We are now ready to prove Theorem \ref{converseweak}. Towards this let $V$ represent the uniform choice of a message in $[2^R]$. Notice that with this choice of $V$ the assumptions of the Theorem \ref{converseweak} imply that 
\beq
\label{sec}
\left\|\rho^{VE}-\rho^V\otimes\rho^E\right\| \leq \delta.
\enq
From \eqref{sec} and setting $\beta\leftarrow\frac{1}{\ln2}$, $\sigma \leftarrow \rho^V \otimes \rho^E$ in Lemma \ref{pinsker} we can now conclude that
\beq
\label{pinsk}
\tr\left[\left\{\rho^{VE} \succ2^{\frac{1}{\ln2}}\rho^V\otimes\rho^E\right\}\rho^{VE}\right] \leq \delta.
\enq
Thus, from \eqref{pinsk} and the definition of smooth max \renyi divergence we can now conclude that 
\beq
\label{mxdivbound}
I^{\delta}_{\infty}\left[V;E\right] \leq 1.5.
\enq
The claim mentioned in Theorem \ref{converseweak} now follows from the following set of inequalities 
\begin{align*}
R &\overset{a}\leq I^{\eps}_{0}\left[V;B\right] \nonumber\\
&\overset{b}\leq \left(I^{\eps}_{0}\left[V;B\right] -I^{\delta}_{\infty}\left[V;E\right]\right)+1.5
\nonumber\\
&\leq \sup_{\left\{V,F\right\}}\left(I^{\eps}_{0}\left[V;B\right] -I^{\delta}_{\infty}\left[V;E\right]\right)+1.5,
\end{align*}
where $a$ follows from \cite[Theorem 1]{wang-renner-prl} and the fact that smooth min \renyi divergence satisfies data processing inequality \cite{wang-renner-prl} and $b$ follows from \eqref{mxdivbound}. This completes the proof of Theorem \ref{converseweak}.

\section{Asymptotics}
In this section we show that the one-shot achievability bounds derived in Theorem \ref{achievability} allow us to characterise the private capacity of the quantum wiretap channel in the information spectrum setting \cite{Hayashi-noniid}. 

Suppose we are given a sequence $\vec{{\bm{\cH}_{BE}}} = \left\{\cH^{(n)}_{BE}\right\}_{n=1}^{\infty}$ of Hilbert spaces and a sequence ${\vec{\cN}} :=\left\{\cN_n\right\}_{n=1}^{\infty}$ of quantum channels where for every $n$, $\cH^{(n)}_{BE}:= \cH^{\otimes n}_{BE}$ and $\cN_n : \mathcal{S}(\cH^{(n)}_A) \to \mathcal{S}(\cH^{(n)}_{BE})$, where $\cH^{(n)}_A=\cH^{\otimes n}_{A}$. We now define achievable rates and capacity for the sequence of wiretap channels ${\vec{\cN}}$.
\begin{definition}
\label{achas}
A rate $R$ is asymptotically achievable for a sequence of quantum wiretap channels ${\vec{\cN}} = \left\{\cN_n\right\}_{n=1}^{\infty}$ if there exists an encoding function $F_{n}$, where $F_{n}: [2^{R^{(n)}}]\to \cS(\cH_A^{\otimes n})$ and decoding POVMs $\{\cT^{B^{(n)}}_{m}: m \in [2^{R^{(n)}}]\}$ such that
$R \leq \liminf_{n \to \infty} \frac{R^{(n)}}{n}$, \\$\lim_{n\to \infty}\frac{1}{2^{R^{(n)}}}\sum_{m \in [2^{R^{(n)}}]}
\tr\left[\left(\mathbb{I} - \cT^B_{m}\right)
      \cN_n(F_n(m))\right]= 0$ and $\lim_{n\to \infty}\left\|\rho^{ME}-\rho^M\otimes \rho^E\right\| = 0.$
 \end{definition}
The supremum of all such achievable rates is called the private capacity of the sequence of quantum wiretap channels ${\vec{\cN}}$ and we represent it by $P({\vec{\cN}})$. 

We now mention some convergence results that would allow us to prove a lower bound on the private capacity of the sequence of wiretap channels ${\vec{\cN}}$. 

\begin{lemma} (Datta and Leditzky \cite{datta-2014-secondorder})
\label{dattacon}
Let ${\vec{\rho}} =\left\{\rho_n\right\}_{n=1}^{\infty}$ and ${\vec{\omega}}= \left\{\omega_n\right\}_{n=1}^{\infty}$ be an arbitrary sequence of states with $\rho_n, \omega_n \in \cS(\cH^{\otimes n})$. Then,
\begin{enumerate}
\item[(i)] $\lim_{\eps \to 0}\liminf_{n \to \infty}\frac{1}{n}{D^{\eps}_{0}\left(\rho_n\|\sigma_n\right)} = \underline{I}\left[{\vec{\rho}};{\vec{\omega}}\right]$
\item[(ii)] $\lim_{\eps \to 0}\limsup_{n \to \infty}\frac{1}{n}{D^{\eps}_{\infty}\left(\rho_n\|\sigma_n\right)}=\overline{I}\left[{\vec{\rho}};{\vec{\omega}}\right]$
\end{enumerate}
\end{lemma}
An immediate consequence of the Theorem \ref{achievability}, Definition \ref{achas} and Lemma \ref{dattacon} is the following corollary.
\begin{corollary}
\label{a}
The private capacity $P({\vec{\cN}})$ of the sequence of quantum wiretap channels ${\vec{\cN}}$ satisfies the following lower bounds
\begin{align*}
P({\vec{\cN}}) \geq \sup_{\{V^n, F_n\}_{n=1}^\infty}\left(\underline{{I}} [\mathbf{V};\mathbf{B}]-\overline{{I}} [\mathbf{V};\mathbf{E}]\right),
\end{align*}
where for every {$n$}, the random variable {$V^n$} takes values over the set
{$\cV^n$}, {$F_n: \cV^n \to \mathcal{S}(\cH_A^{\otimes n})$} and
all the information theoretic quantities are calculated with 
respect to the sequence of state $\vec{\Theta}^{VBE} := \left\{\Theta^{V^nB^nE^n}\right\}_{n=1}^\infty$, where for every $n$
\beq
{\Theta^{V^nB^nE^n}:= \sum_{v^n \in \cV^n}p(v^n)\ket{v^n}\bra{v^n}^{V^n}
\otimes \cN_n^{A^n \to B^n}\left(\rho_{v^n}^{A^{n}}\right)}.\nonumber 
\enq
\end{corollary}
Furthermore, using steps exactly similar to that used in the proof of Theorem \ref{converseweak} we can prove the following corollary.
\begin{corollary}
\label{b}
The private capacity $P({\vec{\cN}})$ of the sequence of quantum wiretap channels ${\vec{\cN}}$ satisfies the following upper bounds
\begin{align*}
P({\vec{\cN}}) \leq \sup_{\{V^n, F_n\}_{n=1}^\infty}\left(\underline{{I}} [\mathbf{V};\mathbf{B}]-\overline{{I}} [\mathbf{V};\mathbf{E}]\right),
\end{align*}
where for every {$n$}, the random variable {$V^n$} takes values over the set
{$\cV^n$}, {$F_n: \cV^n \to \mathcal{S}(\cH_A^{\otimes n})$} and
all the information theoretic quantities are calculated with 
respect to the sequence of state $\vec{\Theta}^{VBE} := \left\{\Theta^{V^nB^nE^n}\right\}_{n=1}^\infty$, where for every $n$
\beq
{\Theta^{V^nB^nE^n}:= \sum_{v^n \in \cV^n}p(v^n)\ket{v^n}\bra{v^n}^{V^n}
\otimes \cN_n^{A^n \to B^n}\left(\rho_{v^n}^{A^{n}}\right)}.\nonumber 
\enq
\end{corollary}
\begin{proof}
Let the rate $R$ be achievable. Therefore, there exists a sequence of codes satisfying the conditions mentioned in Definition \ref{achas}. Furthermore, let $V^n$ be uniformly distributed over $[2^{nR}]$. Notice that for this choice of $V^n$ the conditions mentioned in Definition \ref{achas} imply that 
\beq
\label{independ}
{{\overline{I}}}[{\bm{V}};{\bm{E}}] = 0.
\enq
The claim now follows from the following set of inequalities 
\begin{align}
R &\overset{a}\leq {{\underline{I}}}[{\bf{V}};{\bf{B}}] \nonumber\\
&\overset{b}={{\underline{I}}}[{\bf{V}};{\bf{B}}]-{{\overline{I}}}[{\bf{V}};{\bf{E}}]\nonumber\\
&\leq \max_{\{V_n,F_n\}_{n=1}^{\infty}}\left({{\underline{I}}}[{\bf{V}};{\bf{B}}]-{{\bar{I}}}[{\bf{V}};{\bf{E}}]\right), \nonumber
\end{align}
where $a$ follows from \cite[Lemma 3]{Hayashi-noniid} and $b$ follows from \eqref{independ}. This completes the proof.
\end{proof}
Thus, an immediate consequence of Corollary \ref{a} and Corollary \ref{b} is the following proposition.
\begin{proposition}
The private capacity $P({\vec{\cN}})$ of the sequence of quantum wiretap channels ${\vec{\cN}}$ satisfies the following
\begin{align*}
P({\vec{\cN}}) = \sup_{\{V^n, F_n\}_{n=1}^\infty}\left(\underline{{I}} [\mathbf{V};\mathbf{B}]-\overline{{I}} [\mathbf{V};\mathbf{E}]\right),
\end{align*}
where for every {$n$}, the random variable {$V^n$} takes values over the set
{$\cV^n$}, {$F_n: \cV^n \to \mathcal{S}(\cH_A^{\otimes n})$} and
all the information theoretic quantities are calculated with 
respect to the sequence of states $\vec{\Theta}^{VBE} := \left\{\Theta^{V^nB^nE^n}\right\}_{n=1}^\infty$, where for every $n$
\beq
{\Theta^{V^nB^nE^n}:= \sum_{v^n \in \cV^n}p(v^n)\ket{v^n}\bra{v^n}^{V^n}
\otimes \cN_n^{A^n \to B^n}\left(\rho_{v^n}^{A^{n}}\right)}.\nonumber 
\enq
\end{proposition}

\section{One-Shot Quantum Covering Lemma}
\label{CL}
\begin{theorem}
\label{Cl}
Let $\rv{X}$ be a random variable taking values in the set
$\universe{X}$.  For each $x \in \universe{X}$, let $\rho_x$ be a
quantum state in the space $\hilbertspace{H}$. Let
$\rho = \E_\rv{X}[\rho_\rv{X}]$
be the average of the the states $\rho_x$. Fix $I \geq 0$, and for each $x
\in \universe{X}$ define (based on $I$) the projection $\Pi_x$ and real number
$\eps_x\in \left(0,1\right)$ as follows:
\begin{eqnarray*}
\Pi_x      &=& \{ 2^I \rho \succeq \rho_x\};\\
\eps_x &=&  1 - \tr \left[\Pi_x \rho_x\right].
\end{eqnarray*}
Let $\eps = \E_\rv{X}[\eps_{\rv{X}}]$. Suppose $\rv{s}=(\rv{X}[1],
\rv{X}[2], \ldots, \rv{X}[M])$ is a sequence of independent random
samples drawn according to the distribution of $\rv{X}$, and let
$\tilde{\rho} = \E_{m \in [M]}[\rho_{X[m]}]$.  Then,
\begin{align*}
  \Pr_{\rv{s}}\left\{ \| \tilde{\rho} - \rho \|\geq 22\sqrt{\eps} \right\} \leq 30C\exp\left(-{\frac{10^{-16}\eps^{9}}{\left(\log_2\left(\dim(\cH)\right)\right)^6}}\frac{M}{2^I}\right),
 \end{align*}
 where $C =\dim(\cH)\left(\log_2\left( \frac{4\dim(\cH)}{\eps}\right)+1\right)^2$.
\end{theorem}

\paragraph{Outline:}
To prove this concentration result, we will crucially employ the
Operator Chernoff Bound of Ahlswede and Winter
\cite{ahlswede-02-converse}. We will, however, need to partition the
underlying space so that the operators involved are suitable for an
application of the Operator Chernoff Bound. The operator Chernoff bound requires a lower bound on the the smallest Eigen value of the expectation operator. Hence, we have to partition the space into subspaces such that the ratio of the maximum Eigen value to the minimum Eigen value of expectation operator is not too large. We still have to take care of the very small Eigen values of expectation operator but it turns out that we can simply neglect them introducing only small errors in the process. This strategy of partitioning the space breaks up the operators into blocks. The operator Chernoff bound can directly be applied to the diagonal blocks. However, the off diagonal blocks pose a problem because they are non-Hermitian matrices and non-square matrices and the operator Chernoff bound does not directly apply to them. We handle the off-diagonal blocks separately by proving a new Chernoff bound in terms of the Schatten-infinity norm. We believe the new Chernoff lemma will be of independent interest.

%
First, we present
the tools we will need.

\subsection*{Tools}

\begin{theorem}[Operator Chernoff Bound of Ahlswede and Winter~\cite{ahlswede-02-converse}]
\label{thm:OperatorChernoff}
Let $s=\langle \xi_1, \ldots, \xi_M \rangle$ be $M$ independent and
identically distributed random varaibles taking values as bounded
linear operators in some Hilbert space $\hilbertspace{H}$, such that
$\forall m \in [M]: 0 \preceq \xi_m \preceq \mathbb{I}$. Let the expectation
$\mu = \E[\xi_1]$ satisfy $\mu \succeq a\mathbb{I}$.  Let $\tilde{\xi} =
\E_{m \in [M]} [ \xi_m]$ be the sample average for the above
sample. Then, for $0 < \eta < \frac{1}{2}$, such that $(1+\eta)a \leq
1$, we have 
\beq \Pr\left\{ (1-\eta) \mu \preceq \tilde{\xi} \preceq
(1+\eta) \mu \right\} \geq 1 - 2 \dim(\cH) \exp\left(-\frac{M \eta^2
  a}{2 \ln (2)} \right).
\label{eq:operatorchernoff}
\enq
\end{theorem}

 The bound $\mu \succeq aI$
required in the above theorem will not be naturally available to us
when we apply the above theorem. It will therefore be convenient to state this bound in the following form.
\begin{lemma}
\label{chernoffnewversion}
Let $\rv{X}$ be a random variable taking values in the set
$\universe{X}$.  For each $x \in \universe{X}$, let $\sigma_x$ be a
quantum state in the space $\hilbertspace{H}$ such that $0 \preceq
\sigma_x \preceq \lambda I$. Let $\sigma = \E_\rv{X}[\sigma_\rv{X}]$
be the average of the the states $\sigma_x$. Suppose
$\rv{s}=(\rv{X}[1], \rv{X}[2], \ldots, \rv{X}[M])$ is a sequence of
random samples drawn according to the distribution of $\rv{X}$, and
let $\tilde{\sigma} = \E_{m \in [M]}[\sigma_{\rv{X}[m]}]$.  Let $\eps, \delta > 0 $ be such that
$\eps < \min\{\frac{1}{2}, \frac{\lambda}{\delta}\}$,
then \beq \Pr\left\{\left\|\tilde{\sigma}-\sigma\right\|\leq
\eps\left(\|\sigma\|+\delta\dim(\cH)\right)\right\}\geq
1-2\dim(\cH)\exp\left(\frac{-\eps^2
  M}{2\ln(2)}\frac{\delta}{\lambda+\delta}\right).\label{eq:chversion}
\enq
\end{lemma} 
\begin{proof}
For every $x \in \cX$ we will modify $\sigma_x$ slightly to obtain
$\zeta_x$, and apply Theorem~\ref{thm:OperatorChernoff} to
$\{p_X(x),\zeta_x\}$. Let
$\zeta_x:=(\sigma_x+\delta {I})/(\lambda + \delta)$ and
$\zeta := \mathbb{E}_{\rv{X}}\left[\zeta_\rv{X}\right] =
(\sigma + \delta{I})/(\lambda + \delta)$. Then,
\begin{itemize}
\item $0 \preceq \zeta_x \preceq I$; and
\item $\zeta \succeq  \delta I/(\lambda + \delta)$.
\end{itemize}
We will appy Theorem~\ref{thm:OperatorChernoff} with $\xi_x
\leftarrow \zeta_x$, $\eta \leftarrow \eps$ and $a \leftarrow
\delta/(\lambda + \delta)$. Note that the condition $(1+\eta)a \leq 1$ holds,
since
\[ (1+\eps)\left(\frac{\delta}{\lambda + \delta}\right) 
\leq \left(1 + \frac{\lambda}{\delta}\right)\left(\frac{\delta}{\lambda
  + \delta}\right) = 1.\]
Thus, letting $\tilde{\zeta} = \E_{m \in [M]}[\zeta_{X[m]}]$, we conclude
\begin{align*}
\Pr\left\{\left\|\tilde{\sigma}-\sigma\right\| \leq \eps (\|\sigma\| + \delta \dim(\cH))\right\} &=
\Pr\left\{\left\|\tilde{\zeta}-\zeta\right\| \leq \eps \| \zeta \|
\right\}\\
&\geq 1-2\dim(\cH)\exp\left(\frac{-\eps^2 M}{2\ln\left(2\right)}\frac{\delta}{\lambda+\delta}\right).
\end{align*}
\end{proof}
We will also need a version of the Chernoff bound that is applicable
to rectangular matrices. Versions of the Hoeffding bound and the
Bernstein bound for rectangular matrices (with max-norm instead of
trace-norm) have been derived by Tropp \cite{tropp-2011}. These
results, however, do not seem to be strong enough for our application.
The complete proof of the following version (requiring substantial
work based on Lemma~\ref{chernoffnewversion}) is presented in
Section~\ref{nsq}.

\begin{lemma}
\label{nonsquare}
Let $\rv{X}$ be a random variable taking values in the set
$\universe{X}$.  Let $d_1 \geq d_2$, $\beta \geq 1$, and for each $x
\in \universe{X}$, let $A_x \in \mathbb{C}^{d_1\times d_2}$ such that
$\|A_x\|\leq 1$ and $\|A_x\|_{\infty} \leq \frac{\beta}{d_2}$. Let $A
= \E_\rv{X}[A_\rv{X}]$ be the average of the states $A_x$. Suppose
$\rv{s}=(\rv{X}[1], \rv{X}[2], \ldots, \rv{X}[m])$ is a sequence of
random samples drawn according to the distribution of $X$, and let
$\tilde{A} = \E_{m \in [M]}[A_{\rv{X}[m]}]$.  Then, for $0<\eps <
1$, \beq \Pr_{\rv{s}}\left\{ \| \tilde{A} - A \|\geq \eps
\right\} \leq 25d_1\exp\left( - 10^{-11}{\eps}^3
\frac{M}{\beta}\right). \nonumber
\enq 

\end{lemma}

We will need the Gentle Measurement Lemma of
Winter~\cite[Lemma 9]{winter-99-converse} (see also \cite[Lemma 9.4.2]{wilde-book}).
\begin{lemma} \label{lm:gentlemeasurement}
\label{GMconvex} Let $(p_x, \rho_x)$ be an ensemble of quantum states and 
$\rho = \sum_x p_x \rho_x$.  Let $0 \preceq \Lambda \preceq\mathbb{I}$
and $\eps \in [0,1]$ be such that $\tr[\Lambda\rho] \geq
1-\eps$. Then, \beq
\sum_{x}p_x\left\|\rho_x-\sqrt{\Lambda}\rho_x\sqrt{\Lambda}\right\|
\leq 2 \sqrt{\eps}. \nonumber \enq In particular, if $\Lambda$ is a
projection operator, then $\displaystyle
\sum_{x}p_x\left\|\rho_x-{\Lambda}\rho_x{\Lambda}\right\| \leq 2
\sqrt{\eps}.$
\end{lemma}

\subsection{Proof of Theorem~\ref{Cl}}
As stated above, our proof will be obtained by decomposing the space.
We first present this decomposition.
\paragraph{Decomposition of the space:} We describe the decomposition
by explicitly presenting the orthogonal projector $\Pi_i$ onto the
$i$-th component. Let $D = \dim
\hilbertspace{H}$, and let the spectral decomposition of $\rho$ be
$\sum_{j=1}^D \lambda_j \ketbra{j}$, where $1 \geq \lambda_1 \geq
\lambda_2 \geq \cdots \geq \lambda_D \geq 0$ and $\sum_{j=1}^D
\lambda_j = 1$. 
Then, for $i=1,2,\ldots$, let
\[ \Pi_i = \sum_{j: 2^{-(i-1)} \geq \lambda_j > 2^{-i}}
\ketbra{j}.\]
Let $K= \ceil{\log_2\left( \frac{4\dim(\cH)}{\eps}\right)}$, 
let $\Pi_{\star} = \sum_{i=1}^K \Pi_i$ and 
$\Pi_{\star}^c=\sum_{i> K} \Pi_i$; then 
\begin{equation}
\tr \left[\Pi_{\star}^c \rho\right] \leq \frac{\eps}{4}. \label{eq:epsilon0}
\end{equation}
Thus, intuitively, most of the mass of $\rho$ resides in the subspace
$\Pi_{\star}$. Hence, from triangle inequality and from Lemma \ref{lm:gentlemeasurement} it now follows that to prove Theorem~\ref{Cl} it is sufficient to show that $\|\tilde{\rho} - \Pi^\star \rho \Pi^\star\|$ is small enough.

\paragraph{Proof.}
We will be use the following abbreviations.
\begin{align*}
\rho_x'&:= \Pi_x\rho_x\Pi_x\\
\rho' & := \E_{\rv{X}}[\rho'_{\rv{X}}]\\
\tilde{\rho} &:=\E_{m \in [M]}[\rho_{\rv{X}[m]}]\\
\tilde{\rho}^\prime &:=\E_{ m\in [M]}[\rho'_{\rv{X}[m]}].
\end{align*}
Using the triangle inequality we have the following.
\begin{align}
\left\| \tilde{\rho} - \rho\right\| &\leq
\left\| \tilde{\rho} -  \Pi_{\star}\tilde{\rho}\Pi_{\star} \right\|
+
\left\| {\Pi_{\star}\tilde{\rho}\Pi_{\star}}  - {\Pi_{\star}\tilde{\rho}^\prime\Pi_{\star}}\right\|
+
\left \|{\Pi_{\star}\tilde{\rho}^\prime\Pi_{\star}} - {\Pi_{\star}\rho'\Pi_{\star}} \right\| \nonumber\\
&\hspace{5mm}+
\left \|{\Pi_{\star}\rho'\Pi_{\star}} - {\Pi_{\star}\rho\Pi_{\star}}\right\|
+
\left\| {\Pi_{\star}\rho\Pi_{\star}} - \rho \right\| 
\nonumber\\
&\leq
\left\| \tilde{\rho} -  \Pi_{\star}\tilde{\rho}\Pi_{\star} \right\|
+
\left\| {\tilde{\rho}}  - {\tilde{\rho}^\prime}\right\|
+
\left \|{\Pi_{\star}\tilde{\rho}^\prime\Pi_{\star}} - {\Pi_{\star}\rho'\Pi_{\star}} \right\|
+
\left \|{\rho'} - {\rho}\right\|\nonumber\\
\label{triangterms}
&\hspace{5mm}+
\left\| {\Pi_{\star}\rho\Pi_{\star}} - \rho \right\|. 
\end{align}
The following claims bound the terms on the right.
\begin{claim} We first bound the first and last term.
\label{clai1}
\begin{align}
\label{c1}
\left\| {\Pi_{\star}\rho\Pi_{\star}} - \rho \right\| & \leq \sqrt{\eps};\\
\label{c2}
  \Pr_{\rv{s}}\left\{\left\| \tilde{\rho} -  \Pi_{\star}\tilde{\rho}\Pi_{\star} \right\| \geq \sqrt{\eps}+2\eps\right\} & \leq \exp\left(-2M
{\eps^2}\right).
\end{align}
\end{claim}
\begin{claim} Next, we bound second and the second last term.
\label{clai2}
\begin{align}
\label{c3}
\left \|\rho' - \rho \right\|& \leq 2\sqrt{\eps};\\
\label{c4}
  \Pr_{\rv{s}}\left\{ \left \|\tilde{\rho}  - \tilde{\rho}^\prime\right \|\geq 2\sqrt{\eps}+2\eps\right\} & \leq \exp\left(-2M\eps^2\right).
\end{align}
\end{claim}
\begin{claim} Finally, we bound the middle term.
\label{clai3}
\beq
\label{c5}
\Pr\left\{\left \|{\Pi_{\star}\tilde{\rho}^\prime\Pi_{\star}} - {\Pi_{\star}\rho'\Pi_{\star}}\right\| \geq 8\sqrt{\eps}+4\eps\right\}\leq 28C\exp\left(-{\frac{10^{-16}\eps^{9}}{\left(\log_2\left(\dim(\cH)\right)\right)^6}}\frac{M}{2^I}\right),\enq 
where $C =\dim(\cH)\left(\log_2\left( \frac{4\dim(\cH)}{\eps}\right)+1\right)^2$.
\end{claim}

For now, assume the above mentioned claims (which will proved below),
and observe that from \eqref{triangterms}, \eqref{c1}--\eqref{c5} it
follows that with probability at least $1-
30C\exp\left(-{\frac{10^{-16}\eps^{9}}{\left(\log_2\left(\dim(\cH)\right)\right)^6}}\frac{M}{2^I}\right)$, 
we have, \beq
\label{diff}
\left\| \tilde{\rho} - \rho\right\|\leq 22\sqrt{\eps}.
\enq
We now return to the claims.

\paragraph{\bf{Proof of Claim \ref{clai1}:}} Inequality \eqref{c1} follows immediately
from \eqref{eq:epsilon0} and Lemma~\ref{GMconvex}.  To prove \eqref{c2}
observe that by Lemma~\ref{GMconvex} we have
\begin{align*}
 \mathbb{E}_{\rv{X}}\left[\left\|\rho_{\rv{X}}
-\Pi_{\star}\rho_{\rv{X}}\Pi_{\star}\right\|\right] & \leq \sqrt{\eps}.
\end{align*}
Further, for every $x$, \beq 0\leq\left\|\rho_x
-\Pi_{\star}\rho_x\Pi_{\star}\right\|\leq 2. \nonumber \enq 
Thus, using triangle inequality and the Chernoff bound applied to the
scalar quantitites $\left\|\rho_x
-\Pi_{\star}\rho_x\Pi_{\star}\right\|$, we have
\[
  \Pr_{\rv{s}}\left\{\left\| \tilde{\rho} -  \Pi_{\star}\tilde{\rho}\Pi_{\star} \right\| \geq \sqrt{\eps}+2\eps\right\} \leq  \Pr_{\rv{s}}\left\{
\E_{m \in [M]} \left[\left\|\rho_{\rv{X}[m]}
-\Pi_{\star}\rho_{\rv{X}[m]}\Pi_{\star}\right\|\right]
\geq \sqrt{\eps}+2\eps\right\} 
\leq \exp\left(-2M{\eps^2}\right).
\]

\paragraph{\bf{Proof of Claim \ref{clai2}:}} From our assumption, $\eps =
\E_X[\eps_X]= 1 - \E_{\rv{X}}[\tr\, \Pi_{\rv{X}} \rho_{\rv{X}}]$;
that is, $\E_{\rv{X}}[\tr\, \Pi_{\rv{X}} \rho_{\rv{X}}] =
1-\eps$. Using the triangle inequality,
Lemma~\ref{lm:gentlemeasurement} and Jensen's inequality, we derive \eqref{c3} 
as follows:
\[ \|\rho' - \rho\| \leq \E_{\rv{X}}\left[\|\rho_{\rv{X}}' - \rho_{\rv{X}}\|\right]
                    \leq \E_{\rv{X}}\left[2 \sqrt{\eps_{\rv{X}}} \right]
                    \leq 2 \sqrt{\eps}.\]
Now \eqref{c4} follows by applying the Chernoff bound to the scalar quantities
$\|\rho'_x -\rho_x\|$.

\paragraph{\bf{Proof of Claim \ref{clai3}:}} This claim will need substantial work. Both
expressions $\Pi_{\star}\tilde{\rho}^\prime\Pi_{\star}$ and
$\Pi_{\star}{\rho^\prime}\Pi_{\star}$ involve the operators
$\rho^{\prime}_x=\Pi_x\rho_x\Pi_x$. The first expression is the
expectation of
$\rho_{\star,x}:=\Pi_{\star}{\rho^{\prime}_x}\Pi_{\star}$ over the
samples, whereas the the second expression is the expectation of
$\rho_{\star,x}$ under the original distribution. That these
quantities with high probability should be close to each other would
intuitively follow from the Operator Chernoff Bound. However,
we need to prepare the operators for an application of the bound. In
fact, for every $x \in \cX$ we will obtain a decomposition
\beq
\label{eq:decomposition}
\rho_{\star,x} = \rho^-_{\star,x}+\rho^+_{\star,x}, \enq
($\rho^-_{\star,x}$ and $\rho^+_{\star,x}$ are orthogonal) such that
\beq
\label{decompose}
\mathbb{E}_{\rv{X}}\left[\tr[\rho^+_{\star,\rv{X}}]\right] \leq 4 \eps,
\enq
implying (by Lemma~\ref{lm:gentlemeasurement})
\beq
\label{c31}
\E_X\left[\left\|\rho_{\star,\rv{X}} - \rho^-_{\star,\rv{X}} \right\| \right]  
\leq 4\sqrt{\eps},
\enq
from which by a routine application Chernoff bound to the scalar quantitites
$\left\|\rho_{\star,\rv{X}} - \rho^-_{\star,\rv{X}} \right\|$, it follows that
\beq
\label{c32}
\Pr\left\{\left\|\Pi_{\star}\tilde{\rho}\Pi_{\star}-\E_{m \in[M]}\left[\rho^-_{\star,\rv{X}[m]}\right]\right\|\geq 4\sqrt{\eps}+2\eps\right\} \leq \exp\left(-2\eps^2M\right).
\enq
Using triangle inequality we have
\begin{align}
\label{c33}
\left \|{\Pi_{\star}\tilde{\rho}^\prime\Pi_{\star}} -
      {\Pi_{\star}\rho'\Pi_{\star}} \right\|&\leq
      \left\|{\Pi_{\star}\rho'\Pi_{\star}}-\E_{\rv{X}}\left[\rho^-_{\star,\rv{X}}\right]\right\|
      + \left\|{\Pi_{\star}\tilde{\rho}^\prime\Pi_{\star}}-\E_{m \in
        [M]}\left[\rho^-_{\star,\rv{X}{[m]}}\right]\right\|
      \nonumber\\ &\hspace{3mm}+\left\|\E_{m \in
        [M]}\left[\rho^-_{\star,\rv{X}{[m]}}\right]-\E_{\rv{X}}\left[\rho^-_{\star,\rv{X}}\right]\right\|.
\end{align}
From \eqref{c31} and \eqref{c32} it now follows that with probability at least $1-\exp\left(-2\eps^2M\right)$ we have
\begin{align}
\label{c34}
\left \|{\Pi_{\star}\tilde{\rho}^\prime\Pi_{\star}} - {\Pi_{\star}\rho'\Pi_{\star}} \right\|& \leq 8\sqrt{\eps} + 2 \eps +\left\|\E_{m \in [M]}\left[\rho^-_{\star,\rv{X}{[m]}}\right]-\E_{\rv{X}}\left[\rho^-_{\star,\rv{X}}\right]\right\|.\end{align}
In the following sections, we will establish the following.
\beq
\label{c35}
\Pr\left\{\left\|\E_{m \in
  [M]}\left[\rho^-_{\star,\rv{X}{[m]}}\right]-\E_{\rv{X}}\left[\rho^-_{\star,\rv{X}}\right]\right\|\geq
2\eps\right\} \leq
27C\exp\left(-{\frac{10^{-16}\eps^{9}}{\left(\log_2\left(\dim(\cH)\right)\right)^6}}\frac{M}{2^I}\right),\enq 
where $C =\dim(\cH)\left(\log_2\left( \frac{4\dim(\cH)}{\eps}\right)+1\right)^2$.
Claim \ref{clai3} thus follow from
\eqref{c31}, \eqref{c32}, \eqref{c33}, \eqref{c34} and \eqref{c35}. To
complete the proof, we need to provide the decomposition stated in
\eqref{eq:decomposition} and establish \eqref{decompose} and \eqref{c35}. The
subsequent sections will be devoted to these.

\subsection{Decomposition of $\rho_{\star,x}$ as $\rho^+_{\star,x}+\rho^-_{\star,x}$}
Recall that 
$$\rho_{\star,x} =\Pi_{\star}\Pi_x\rho_x\Pi_x\Pi_{\star},$$
where $\Pi_{\star}=\sum_{i=1}^K\Pi_i$. For $i\in\{1,2,\cdots,K\}$ and $x\in\cX$, let
\beq
\Pi^{+}_{i,x} := \{\Pi_{i}\Pi_{x}\rho\Pi_{x}\Pi_i \succ 4\Pi_i\rho\Pi_i\}\nonumber
\enq
and $\Pi^{-}_{i,x}:=\Pi_i-\Pi^+_{i,x}$. Letting $\Pi^{-}_{\star,x}:=\sum_{i=1}^K\Pi^{-}_{i,x}$ and $\Pi^+_{\star,x}:=\Pi_\star-\Pi^{-}_{\star,x}$ we now have the following decomposition of $\rho_{\star,x}$
\beq
\rho_{\star,x} =
\Pi^{-}_{\star,x}\rho_{\star,x}\Pi^{-}_{\star,x}+\Pi^{+}_{\star,x}\rho_{\star,x}\Pi^{+}_{\star,x}, 
\enq
and, we have, in particular, letting $\lambda_{\max}(\Pi_i \rho\Pi_i)$ the maximum Eigen value of the operator $\Pi\rho \Pi_i$, we have the following set of inequalities
\begin{align}
\Pi_{i,x}^{-} \Pi_x \rho_x \Pi_x \Pi_{i,x}^{-} & \overset{a}\preceq 2^I \Pi_{i,x}^{-} \Pi_x \rho \Pi_x \Pi_{i,x}^{-} \nonumber\\
\label{eq:piminusproperty}
& \overset{b}\preceq 2^{I+2} \lambda_{\max}(\Pi_i \rho\Pi_i) \Pi_i,
\end{align}
where $a$ follows from the fact that $\Pi_x \rho_x \Pi_x \preceq 2^I \Pi_x\rho\Pi_x$ and $b$ follows from the definition of $\Pi_{i,x}^{-}$ and the fact that $\Pi_{i,x}^{-} \preceq \Pi_i$
\subsubsection{Proof of \eqref{decompose}}
We will need the following key lemma.
\begin{lemma}
\label{keylemma}
\begin{description}
\item[(a)] If $\bra{v}\Pi_x\rho\Pi_x\ket{v} > 4 \bra{v}\rho\ket{v}$ then $\bra{v}\Pi_x\rho\Pi_x\ket{v}<4\bra{v}\Pi^c_x\rho\Pi^c_x\ket{v}$.
\item[(b)] Let $\Pi$ be some projection and $\Pi^+_{x}$ be its subspace defined by $\Pi^+_x:=\left\{\Pi\Pi_x\rho\Pi_x\Pi\succ4\Pi\rho\Pi\right\}$. Then, $\tr\left[\Pi^+_x\Pi_x\rho\Pi_x\Pi^+_x\right]\leq 4\tr\left[\Pi^+_x\Pi^c_x\rho\Pi^c_x\Pi^+_x\right] $.
\end{description}
\end{lemma}
\begin{proof}
\begin{description}
\item[(a)] We have 
\begin{align}
\bra{v}\Pi_x\rho\Pi_x\ket{v} &>4 \bra{v}\rho\ket{v}\nonumber\\
\label{rearr}
&=4\left(\bra{v}\Pi_x\rho\Pi_x\ket{v}+\bra{v}\Pi_x\rho\Pi^c_x\ket{v}+\bra{v}\Pi^c_x\rho\Pi_x\ket{v}+\bra{v}\Pi^c_x\rho\Pi^c_x\ket{v}\right)
\end{align}
By rearranging the terms of \eqref{rearr} we get 
\begin{align*}
0 & > 3\bra{v}\Pi_x\rho\Pi_x\ket{v} + 4\bra{v}\Pi^c_x\rho\Pi^c_x\ket{v}
      + 4\left(\bra{v}\Pi_x\rho\Pi^c_x\ket{v}+\bra{v}\Pi^c_x\rho\Pi_x\ket{v}\right)\\
& \geq  3\bra{v}\Pi_x\rho\Pi_x\ket{v} + 4\bra{v}\Pi^c_x\rho\Pi^c_x\ket{v}
     - 4 \left|\bra{v}\Pi_x\rho\Pi^c_x\ket{v}+\bra{v}\Pi^c_x\rho\Pi_x\ket{v}\right|\\
& \geq  3\bra{v}\Pi_x\rho\Pi_x\ket{v} + 4\bra{v}\Pi^c_x\rho\Pi^c_x\ket{v}
     - 4 \left|\bra{v}\Pi_x\rho\Pi^c_x\ket{v}\right|  - 4
         \left|\bra{v}\Pi^c_x\rho\Pi_x\ket{v}\right|\\
& \geq  3\bra{v}\Pi_x\rho\Pi_x\ket{v} + 4\bra{v}\Pi^c_x\rho\Pi^c_x\ket{v}
- 4\big|\bra{v}\Pi_x\rho^{\frac{1}{2}}\rho^{\frac{1}{2}}\Pi^c_x\ket{v}\big|
-4\big|\bra{v}\Pi^c_x\rho^{\frac{1}{2}}\rho^{\frac{1}{2}}\Pi_x\ket{v}\big| \\
&\overset{a} \geq  3\bra{v}\Pi_x\rho\Pi_x\ket{v} + 4\bra{v}\Pi^c_x\rho\Pi^c_x\ket{v}
- 8\sqrt{\bra{v}\Pi_x\rho\Pi_x\ket{v}\bra{v}\Pi^c_x\rho\Pi^c_x\ket{v}}\\
& =
\left(3\sqrt{\alpha}-2\sqrt{\beta}\right)\left(\sqrt{\alpha}-2\sqrt{\beta}\right)\\
&= 3\left(\sqrt{\alpha} - \frac{2}{3} \sqrt{\beta}\right)\left(\sqrt{\alpha}
                - 2 \sqrt{\beta}\right)
,
\end{align*}
where $a$ follows from the Cauchy-Schwarz inequality,
$\alpha=\bra{v}\Pi_x\rho\Pi_x\ket{v}$ and
$\beta=\bra{v}\Pi^c_x\rho\Pi^c_x\ket{v}$.
Thus, $\frac{2}{3}\sqrt{\beta}< \sqrt{\alpha} < 2\sqrt{\beta}$. 
In particular, 
$\bra{v}\Pi_x\rho\Pi_x\ket{v}
= \alpha< 4\beta = 4\bra{v}\Pi^c_x\rho\Pi^c_x\ket{v}$.
\item[(b)] We will use part (a). Let $\Pi^+_x=\sum_{i=1}^d\ket{v_i}\bra{v_i}$, where $\left\{\ket{v_i}\right\}$ is an orthonormal basis for $\Pi^+_x$. Then,
\begin{align}
\tr\left[\Pi^+_x\Pi_x\rho\Pi_x\Pi^+_x\right]&=\sum_{i=1}^d\bra{v_i}\Pi_x\rho\Pi_x\ket{v_i}\nonumber\\
&\leq4\sum_{i=1}^d\bra{v_i}\Pi^c_x\rho\Pi^c_x\ket{v_i}\nonumber\\
&=\tr\left[\Pi^+_x\Pi^c_x\rho\Pi^c_x\Pi^+_x\right]
\end{align}
where the inequality above follows from part (a) and the definition of
$\Pi^+_x$.
\end{description}
\end{proof}

We now have all the ingredients to prove \eqref{decompose}. 
\begin{align}
\tr\left[\Pi^+_{\star,x}\rho_{\star,x}\Pi^+_{\star,x}\right] &\overset{a}=\sum_{i=1}^K\tr\left[\Pi^+_{i,x}\rho_{\star,x}\Pi^+_{i,x}\right]\nonumber\\
&\overset{b}=\sum_{i=1}^K\tr\left[\Pi^+_{i,x}\Pi_x\rho_{x}\Pi_x\Pi^+_{i,x}\right]\nonumber\\
&\overset{c}\leq2^I\sum_{i=1}^K\tr\left[\Pi^+_{i,x}\Pi_x\rho\Pi_x\Pi^+_{i,x}\right]\nonumber\\
&\overset{d}\leq 2^I\sum_{i=1}^K4\tr\left[\Pi^+_{i,x}\Pi^c_x\rho\Pi^c_x\Pi^+_{i,x}\right]\nonumber\\
&\overset{e}\leq 4\sum_{i=1}^K\tr\left[\Pi^+_{i,x}\Pi^c_x\rho_x\Pi^c_x\Pi^+_{i,x}\right]\nonumber\\
\label{exptr}
&\overset{f}\leq  4 \tr\left[\Pi^c_x\rho_x\Pi^c_x\right],
\end{align}
where $a$ follows from the definition $\Pi^+_{\star,x}$ and from the circular property of the trace and the fact that $\Pi^+_{i,x}\Pi^+_{j,x}= 0,$ for $i \neq j$; $b$ follows
from the fact that $\Pi^+_{i,x}\preceq\Pi_{\star}$; $c$ follows from
the definition of $\Pi_x$; $d$ follows from the definition of
$\Pi^+_{i,x}$ and Lemma \ref{keylemma}; $e$ follows from the
definition of $\Pi^c_x$ and $f$ follows because
$\sum_{i=1}^K\Pi^+_{i,x}\preceq \mathbb{I}$.Thus, \eqref{decompose}
now follows by computing the expectation of both sides of \eqref{exptr},
using our assumption $\tr \Pi^c_x \rho_x \leq \eps_x$ and
$\E_\rv{x}[\eps_\rv{x}] = \eps$.

\subsubsection{Proof of \eqref{c35}}
We will use the Operator Chernoff bound as stated in 
Lemma~\ref{chernoffnewversion}. 
\begin{claim}
\label{diagclaim}
For $i=1,2,\ldots,K$, we have
\begin{align}
\Pr&\left\{\left\|\E_{m \in
  [M]}[\Pi^{-}_{i,\rv{X}[m]}\rho'_{\rv{X}[m]}\Pi^{-}_{i,\rv{X}[m]}]-\E_{\rv{X}}[\Pi^{-}_{i,\rv{X}}\rho'_{\rv{X}}\Pi^{-}_{i,\rv{X}}]\right\|\leq\frac{\eps}{2}\left(\|\mu_i\|+ \|\Pi_i\rho \Pi_i\|\right)\right\} \nonumber \\
  &\geq1-2\dim(\Pi_i)\exp\left(\frac{-\eps^2}{8\ln2}\frac{M}{2^{I+3}+1}\right),
\label{eq:cl4}
\end{align}
where $\mu_i=\E_{\rv{X}}[\Pi^{-}_{i,\rv{X}}\rho'_{\rv{X}}\Pi^{-}_{i,\rv{X}}]$.
\end{claim}
We appy Lemma~\ref{chernoffnewversion} with $\cH \leftarrow \Pi_i$;
$\sigma_{x}\leftarrow\Pi^{-}_{i,x}\rho'_{x}\Pi^{-}_{i,x}$;
$\eps\leftarrow\frac{\eps}{2}$ and
$\delta\leftarrow\lambda_{\min}(\Pi_i\rho\Pi_i)$.
Then, $\delta \cdot \dim(\cH) = \lambda_{\min}(\Pi_i\rho\Pi_i)\cdot
\dim(\cH) \leq \| \Pi_i \rho \Pi_i\|$.  With this the LHS of
(\ref{eq:chversion}) matches the LHS of (\ref{eq:cl4}). 

Next consider the RHS. 
Recall from (\ref{eq:piminusproperty}) that 
$\Pi_{i,x}^{-} \Pi_x \rho_x \Pi_x \Pi_{i,x}^{-} \preceq 2^{I+2} \lambda_{\max}(\Pi_i \rho\Pi_i) \Pi_i
$. Thus, $0 \preceq \sigma_x \preceq 2^{I+2} \lambda_{\max} (\Pi_i \rho \Pi_i)\Pi_i$.
Now, $2^{I+2}
\lambda_{\max}(\Pi_i \rho \Pi_i) \leq 2^{I+3} \lambda_{\min}(\Pi_i \rho
\Pi_i) = 2^{I+3} \delta$.  So, under our substitution
the RHS of (\ref{eq:chversion}) is at least the RHS of (\ref{eq:cl4}).
Thus (\ref{eq:cl4}) is justified.

We are now ready to establish \eqref{c35}. We will account for the contributions from the diagonal and off-diagonal blocks separately. We have
from the definition of $\Pi^{-}_{\star, x}$ and triangle inequality that
\begin{align}
\left\|\E_{m \in [M]}\left[\rho^-_{\star,\rv{X}{[m]}}\right]-\E_{\rv{X}}\left[\rho^-_{\star,\rv{X}}\right]\right\|& \leq \sum_{i=1}^{K}\left\|\E_{m \in
  [M]}[\Pi^{-}_{i,\rv{X}[m]}\rho'_{\rv{X}[m]}\Pi^{-}_{i,\rv{X}[m]}]-\E_{\rv{X}}[\Pi^{-}_{i,\rv{X}}\rho'_{\rv{X}}\Pi^{-}_{i,\rv{X}}]\right\|\nonumber\\
\label{piminusterms}
 &\hspace{3.5mm}+ \sum_{i\neq l}\left\|\E_{m \in [M]}[\Pi^{-}_{i,\rv{X}[m]}\rho'_{\rv{X}[m]}\Pi^{-}_{l,\rv{X}[m]}]-\E_{\rv{X}}[\Pi^{-}_{i,\rv{X}}\rho'_{\rv{X}}\Pi^{-}_{l,\rv{X}}]\right\|
\end{align}

\paragraph{The diagonal blocks:} 
From Claim \ref{diagclaim} and union bound it follows that 
with probability at least $1-2\dim(\cH)$ $\exp\left(\frac{-\eps^2}{8\ln2}\frac{M}{2^{I+3}+1}\right)$ we have
\begin{align}
\sum_{i=1}^{K}\left\|\E_{m \in
  [M]}[\Pi^{-}_{i,\rv{X}[m]}\rho'_{\rv{X}[m]}\Pi^{-}_{i,\rv{X}[m]}]-\E_{\rv{X}}[\Pi^{-}_{i,\rv{X}}\rho'_{\rv{X}}\Pi^{-}_{i,\rv{X}}]\right\|&\leq \frac{\eps}{2}\sum_{i=1}^{K}\|\mu_i\|+\frac{\eps}{2}\sum_{i=1}^{K}\|\Pi_i\rho\Pi_i\|\nonumber\\
  \label{diagminus}
  &\leq\eps.
\end{align}

\paragraph{Off-diagonal blocks:} For every $i, l \in [1:K]$ ($i \neq l$) it follows by $\eps \leftarrow \frac{\eps}{K^2}$ in Theorem \ref{nondiagtheo} that
\begin{align}
\Pr&\left\{\left\|\E_{m \in [M]}[\Pi^{-}_{i,\rv{X}[m]}\rho'_{\rv{X}[m]}\Pi^{-}_{l,\rv{X}[m]}]-\E_{\rv{X}}[\Pi^{-}_{i,\rv{X}}\rho'_{\rv{X}}\Pi^{-}_{l,\rv{X}}]\right\|\geq \frac{\eps}{K^2} \right\} \nonumber\\
\label{errnondiago}
& \geq 1- 25\dim(\cH)\exp\left(-{10^{-12}\frac{\eps^{3}}{K^6}}\frac{M}{2^I}\right)
\end{align}
From \eqref{errnondiago} and letting $K = \ceil{\log_2\left( \frac{4\dim(\cH)}{\eps}\right)}$ it now follows using union bound that with probability at least $1-25\dim(\cH)\left(\log_2\left( \frac{4\dim(\cH)}{\eps}\right)+1\right)^2\exp\left(-{\frac{10^{-16}\eps^{9}}{\left(\log_2\left(\dim(\cH)\right)\right)^6}}\frac{M}{2^I}\right)
$, we have
\beq
\label{errnondiago12} 
\sum_{i\neq l}\left\|\E_{m \in [M]}[\Pi^{-}_{i,\rv{X}[m]}\rho'_{\rv{X}[m]}\Pi^{-}_{l,\rv{X}[m]}]-\E_{\rv{X}}[\Pi^{-}_{i,\rv{X}}\rho'_{\rv{X}}\Pi^{-}_{l,\rv{X}}]\right\| \leq \eps. 
\enq
Thus, from \eqref{piminusterms}, \eqref{diagminus} and \eqref{errnondiago12} and union bound it follows that
\beq
\left\|\E_{m \in [M]}\left[\rho^-_{\star,\rv{X}{[m]}}\right]-\E_{\rv{X}}\left[\rho^-_{\star,\rv{X}}\right]\right\| \leq 2\eps, \nonumber
\enq
with probability at least $27\dim(\cH)\left(\log_2\left( \frac{4\dim(\cH)}{\eps}\right)+1\right)^2\exp\left(-{\frac{10^{-16}\eps^{9}}{\left(\log_2\left(\dim(\cH)\right)\right)^6}}\frac{M}{2^I}\right)$.

\begin{theorem}
\label{nondiagtheo}
Let $\rv{X}$ be a random variable taking values in a set
$\universe{X}$.  For each $x \in \universe{X}$, let $\rho_x$ be a
quantum state in the space $\hilbertspace{H}$. Let
$\rho = \E_X[\rho_X]$
be the average of the the states $\rho_x$. Furthermore, let $\Pi_i=\sum_{j: 2^{-(i-1)} \geq \lambda_j > 2^{-i}}
\ketbra{j}$ and $\Pi_l=\sum_{j: 2^{-(l-1)} \geq \lambda_j > 2^{-l}}
\ketbra{j}$, where for every $j$, $\ket{j}$ and $\lambda_j$ represent the Eigen vector and the corresponding Eigen value of $\rho$ and $i\neq l$. Fix $I >0$, and for each $x
\in \universe{X}$ define (based on $I$) the projections $\Pi_x$, $\Pi^{-}_{i,x}, \Pi^{-}_{l,x}$, operators $\rho^{\prime}_x, \sigma^{-}_{i,l,x}$ as follows:
\begin{eqnarray}
\label{pix}
\Pi_x      &=& \{ 2^I \rho \succeq \rho_x\};\\
\rho^{\prime}_x &=&\Pi_x\rho_x\Pi_x;\\
\label{Pixminus1}
\Pi^{-}_{i,x} &= &\{\Pi_{i}\Pi_{x}\rho\Pi_{x}\Pi_i \preceq 4\Pi_i\rho\Pi_i\}\\
\label{Pixminus2}
\Pi^{-}_{l,x} &= &\{\Pi_{l}\Pi_{x}\rho\Pi_{x}\Pi_l \preceq 4\Pi_l\rho\Pi_l\}\\
\label{defsigmaminusi,l,x}
\sigma^{-}_{i,l, x} &=& \Pi^-_{i,x}\rho^{\prime }_{x}\Pi^{-}_{l,x}
\end{eqnarray}
Let $\rv{s}=(\rv{X}[1], \rv{X}[2], \ldots,
\rv{X}[M])$ be a sequence of $M$ random samples drawn according to the
distribution of $\rv{X}$, and let $\tilde{\sigma}_{i,l}^- = \E_{m \in
  [M]}[\sigma^{-}_{i,l,\rv{X}[m]}]$ and $\sigma^{-}_{i,l}=\E_{X}\left[\sigma^{-}_{i,l,\rv{X}}\right]$.
Then, for $0 < \eps < 1$,
\begin{align}
\Pr_{\rv{s}}\left\{ \| \tilde{\sigma}_{i,l}^- - \sigma^{-}_{i,l}  \| 
     \geq 
 \eps\right\} \leq 25\dim(\cH)\exp\left(-{10^{-12}\eps^{3}}\frac{M}{2^I}\right).\nonumber
\end{align}
\end{theorem}

\begin{proof}
Let $\lambda_{\min}(i)$ and $\lambda_{\max}(i)$ be the minimum and maximum Eigen values of the operator $\Pi_i\rho\Pi_i$ and analogously $\lambda_{\min}(l)$ and $\lambda_{\max}(l)$ represent the minimum and maximum Eigen values of the operator $\Pi_l\rho\Pi_l$.  For each $x \in \cX$, notice the following set of inequalities
\begin{align}
\left\|\sigma^{-}_{i,l,x}\right\|_{\infty} &= \left\|\Pi^-_{i,x}\Pi_{x}\rho_{x}\Pi_{x}\Pi^{-}_{l,x}\right\|_{\infty}\nonumber\\
&\overset{a}=\max_{u,w:\|u\|,\|w\|=1}{\left|\bra{u}\Pi^-_{i,x}\Pi_{x}\rho_{x}\Pi_{x}\Pi^{-}_{l,x}\ket{w}\right|} \nonumber\\
&\overset{b} \leq \sqrt{\left|\bra{u}\Pi^-_{i,x}\Pi_{x}\rho_{x}\Pi_{x}\Pi^{-}_{i,x}\ket{u}\right|}\sqrt{\left|\bra{w}\Pi^-_{l,x}\Pi_{x}\rho_{x}\Pi_{x}\Pi^{-}_{l,x}\ket{w}\right|} \nonumber\\
&\overset{c}\leq \sqrt{\left\|\Pi^-_{i,x}\Pi_{x}\rho_{x}\Pi_{x}\Pi^{-}_{i,x}\right\|_{\infty}}\sqrt{\left\|\Pi^-_{l,x}\Pi_{x}\rho_{x}\Pi_{x}\Pi^{-}_{l,x}\right\|_{\infty}}\nonumber\\
\label{binfi1}
&\overset{d}\leq 2^{I+3}\sqrt{\lambda_{\min}(i)\lambda_{\min}(l)},
\end{align}
where $a$ follows from the definition of the the infinity norm, $b$ follows from Cauchy Schwarz inequality, $c$ again follows from the definition of infinity norm and $d$ follows because of the following set of inequalities
\begin{align}
\Pi^-_{i,x}\Pi_{x}\rho_{x}\Pi_{x}\Pi^{-}_{i,x}&\overset{a}\preceq 2^{I}\Pi^-_{i,x}\Pi_{x}\rho\Pi_{x}\Pi^{-}_{i,x}\nonumber\\
&\overset{b}\preceq2^{I+2}\Pi^-_{i,x}\rho\Pi^{-}_{i,x}\nonumber\\
\label{normi1}
&\overset{c}= 2^{I+2}\Pi^-_{i,x}\Pi_i\rho\Pi_i\Pi^{-}_{i,x},
\end{align}
where $a$ follows from the definition of $\Pi_x$ \eqref{pix}, $b$ follows from the definition of $\Pi^{-}_{i,x}$ \eqref{Pixminus1}, $c$ follows because the projector $\Pi^-_{i,x}$ projects onto a subspace of $\Pi_i$. Thus, from \eqref{normi1} it now follows that 
\begin{align}
\left\|\Pi^-_{i,x}\Pi_{x}\rho_{x}\Pi_{x}\Pi^{-}_{i,x}\right\|_{\infty}& \leq 2^{I+2}\left\|\Pi^-_{i,x}\Pi_i\rho\Pi_i\Pi^{-}_{i,x}\right\|_{\infty}\nonumber\\
&\overset{a}\leq 2^{I+2}\left\|\Pi_i\rho\Pi_i\right\|_{\infty}\nonumber\\
&\leq 2^{\left(I+2\right)}\lambda_{\max}(i)\nonumber\\
&\overset{b}\leq 2^{I+3}\lambda_{\min}(i),
\end{align}
where $a$ follows from \cite[Proposition IV.2.4]{bhatia-1997} and $b$ follows from the fact that $\frac{\lambda_{\max}(i)}{\lambda_{\min}(i)}\leq 2$. Similarly, $\left\|\Pi^-_{l,x}\Pi_{x}\rho_{x}\Pi_{x}\Pi^{-}_{l,x}\right\|_{\infty} \leq 2^{I+3}\lambda_{\min}(l)$. Also, from the fact that $\left\|\Pi_{x}\rho_{x}\Pi_{x}\right\| \leq 1$ it follows that 
\beq
\label{tracenormnonsquare1}
\left\|\sigma^{-}_{i,l,x}\right\|\leq1. 
\enq
For each $x \in \cX$, let us view the operator $\sigma^{-}_{i,j,x}$ as non-square matrix in $\mathbb{C}^{\dim(\Pi_i)\times \dim(\Pi_l)}$ embedded inside $\mathbb{C}^{\dim(\cH)\times\dim(\cH)}$ matrix. It now follows from \eqref{binfi1} and \eqref{tracenormnonsquare1} that both the conditions required for the application of Lemma \ref{nonsquare} are satisfied. The claim now immediately follows by $\beta \leftarrow 2^{I+3}$ in the Lemma \ref {nonsquare}.
This completes the proof.
\end{proof}
\section{Chernoff Bound for non-square matrices}
\label{nsq}
In this section, we prove a concentration result for non-square
matrices which need not be positive. We first restrict attention 
to square (but not necessarily positive) matrices.
\begin{lemma}
\label{nonpositive}
Let $\rv{X}$ be a random variable taking values in a set
$\universe{X}$.  For each $x \in \universe{X}$, let $A_x \in
\mathbb{C}^{d\times d}$ be a (not necessarily positive) matrix. Let
$\mu \geq 0$ and $, \beta\geq 1$ be such that $\|A_x\| \leq \mu $ and
$\|A_x\|_{\infty} \leq \frac{\beta}{d}$ for all $x \in
\universe{X}$. Let $A=\mathbb{E}_{\rv{X}}\left[A_{\rv{X}}\right]$ be
the average of the matrices $A_x$. Suppose $\rv{s}=(\rv{X}[1],
\rv{X}[2], \ldots, \rv{X}[M])$ is a sequence of random samples drawn
according to the distribution of $X$, and  $\tilde{A} = \E_{m \in
  [M]}[A_{\rv{X}[m]}]$.  Then, for $0 < \eps < \frac{1}{2}$,
\beq \Pr_{\rv{s}}\left\{ \| \tilde{A} -
A \|\leq \eps \right\}\geq
1-4d\exp\left(\frac{-\eps^2}{32\ln(2)\mu}\frac{M}{{2\beta}+\mu}\right).
\enq
\end{lemma}
\begin{proof}
We will establish our claim by embedding each $A_x$ in a matrix let
$B_x \in \mathbb{C}^{2d\times 2d}$ as follows.  Let
$A_x=\sum_{k=1}^d\lambda_k\ket{v_k}\bra{w_k}$ where $\{v_k\}_{k=1}^d$
and $\{w_k\}_{k=1}^d$ are orthonormal bases and $\lambda_k \geq 0$.  We
enlarge the Hilbert space to obtain
$\cH^\prime=\mathbb{C}^2\otimes\mathbb{C}^d$ of dimension $2d$, where
we view $\mathbb{C}^2$ as a space of single qubit. Then, for each
$v_k$ let $\ket{\tilde{v}_k}=\ket{0}\ket{v_k}$ and similarly for each
$w_k$ let $\ket{\tilde{w}_k}=\ket{1}\ket{w_k}$; then, the set
$\left\{\ket{\tilde{v}_k}:k=1,\cdots,d\right\}\cup\left\{\ket{\tilde{w}_k}:k=1,\cdots,d\right\}$
is an orthonormal basis for $\cH^\prime$. Let \beq B_x = \sum_{k=1}^d
\lambda_k\left(\ket{\tilde{v}_k+\tilde{w}_k}\bra{\tilde{v}_k+\tilde{w}_k}\right).
\enq 
Clearly $B_x$ is a positive operator with the following spectral
decomposition. 
\beq
\label{specB_x}
B_x=\sum_{k=1}^d2\lambda_k
\left(\frac{\ket{\tilde{v}_k+\tilde{w}_k}}{\sqrt{2}}\right)\left(
\frac{\bra{\tilde{v}_k+\tilde{w}_k}}{\sqrt{2}}\right)+\sum_{k=1}^d0\left(\frac{\ket{\tilde{v}_k-\tilde{w}_k}}{\sqrt{2}}\right)\left(\frac{\bra{\tilde{v}_k-\tilde{w}_k}}{\sqrt{2}}\right).
\enq 

In particular, we have $\left\|B_x\right\|\leq2\left\|A_x\right\|$ and
$\left\|B_x\right\|_{\infty}\leq2\left\|A_x\right\|_{\infty}$.  We
will apply Lemma~\ref{chernoffnewversion} to obtain a concentration
with for the matrices $B_x$, and argue that a similar concentration
must hold for $A_x$.  Let $\tilde{B}=\E_{m \in [M]}[B_{\rv{X}[m]}]$
and $B=\mathbb{E}_{\rv{X}}\left[B_{\rv{X}}\right]$.  We have $ 0
\preceq B_x \preceq \frac{2\beta}{d}\mathbb{I}$, $\|B_x\| \leq 2 \mu$
and $\|B\| \leq 2\mu$.  We apply Lemma~\ref{chernoffnewversion} with
$\sigma_x \leftarrow B_x$, $\lambda \leftarrow \frac{2\beta}{d}$,
$\eps \leftarrow \frac{\eps}{4\mu}$ and $\delta \leftarrow
\frac{\mu}{d}$ (note that $\frac{\lambda}{\delta} =
\frac{2\beta}{\mu}> \frac{\eps}{4\mu}$) and conclude that
\begin{align}
\Pr\left\{\left\|\tilde{B}-B\right\|\leq
\frac{\eps}{4\mu}\left(2\mu+\delta2d\right)\right\}&\geq
1-4d\exp\left(\frac{-\eps^2
  M}{32\ln(2)\mu^2}\frac{\frac{\mu}{d}}{\frac{2\beta}{d}+\frac{\mu}{d}}\right)\nonumber\\
  & = 1-4d\exp\left(\frac{-\eps^2}{32\ln(2)\mu}\frac{M}{{2\beta}+\mu}\right).\nonumber 
\end{align}
Notice that the operator $B_x$ has the following form.
\beq
\label{intermediary22}
B_{x}=
\left[\begin{array}{c|c}
 &\\
 \sum_{k=1}^{d}\lambda_k\ket{\tilde{v}_k}\bra{\tilde{v}_k} & A_x\\
  &\\\hline
  &\\
 A^\dagger_x & \sum_{k=1}^{d}\lambda_k\ket{\tilde{w}_k}\bra{\tilde{w}_k}\\
 &\\
\end{array}\right].
\enq
Thus,
\begin{align*}
\Pr\left\{\left\|\tilde{A}-A\right\|\leq \eps\right\}&\geq \Pr\left\{\left\|\tilde{B}-B\right\|\leq \eps\right\}
\geq 1-4d\exp\left(\frac{-\eps^2}{32\ln(2)\mu}\frac{M}{{2\beta}+\mu}\right).
\end{align*}
\end{proof}

\begin{corollary}\label{cor:tononpositive}
Let $\rv{X}$ be a random variable taking values in the set
$\universe{X}$.  For each $x \in \universe{X}$, let $A_x \in
\mathbb{C}^{d_1\times d_2}$ ($d_2 \leq d_1 \leq 2d_2$) be a
non-positive matrix together with the property that $\|A_x\| \leq \mu
$ and $\|A_x\|_{\infty} \leq \frac{\beta}{d_1}$. Let
$A=\mathbb{E}_{\rv{X}}\left[A_{\rv{X}}\right]$ be the average of the
matrices $A_x$ Suppose $\rv{s}=(\rv{X}[1], \rv{X}[2], \ldots,
\rv{X}[M])$ be a sequence of random samples drawn according to the
distribution of $X$, and let $\tilde{A} = \E_{m \in
  [M]}[A_{\rv{X}[m]}]$.  Then, for $0<\eps<\frac{1}{2}$
\beq 
 \Pr_{\rv{s}}\left\{ \| \tilde{A} -
A \|\leq \eps) \right\}\leq
1-4d_1\exp\left(\frac{-\eps^2}{32\ln(2)\mu}\frac{M}{{2\beta}+\mu}\right).
\enq
\end{corollary}
\begin{proof}
This claim differs from Lemma~\ref{nonpositive} because we do not 
require $A_x$ to be a square matrix.  We can obtain a matrix
$B_x$ from $A_x$ by adding $d_1-d_2$ all zeroes columns.  Let the
corresponding expectations be $B=
\mathbb{E}_{\rv{X}}\left[B_{\rv{X}}\right]$ and $\tilde{B}=\E_{m \in
  [M]}[B_{\rv{X}[m]}]$.  Note that $\|B_x\| = \|A_x\| \leq \mu$,
$\|B_x\|_\infty=\|A_x\|_\infty \leq \frac{\beta}{d_1}$ and $\|\tilde{B} -
B\|= \|\tilde{A} - A\|$. The claim then follows immediately by taking
$A_x \leftarrow B_x$ in the Lemma~\ref{nonpositive}.
\end{proof}

\subsection{Concentration result for non-square matrices}
We can now show the main concentration result of this section.
\begin{lemma}
\label{nonsquare}
Let $\rv{X}$ be a random variable taking values in the set
$\universe{X}$.  For each $x \in \universe{X}$, let $A_x \in
\mathbb{C}^{d_1\times d_2}$ such that $d_1 \geq d_2$, $\|A_x\|\leq 1$
and $\|A_x\|_{\infty} \leq \frac{\beta}{d_2}$, where $\beta \geq 1$. Let $A =
\E_\rv{X}[A_\rv{X}]$ be the average of the states $A_x$. Suppose
$\rv{s}=(\rv{X}[1], \rv{X}[2], \ldots, \rv{X}[m])$ be a sequence of
random samples drawn according to the distribution of $X$, and let
$\tilde{A} = \E_{m \in [M]}[A_{\rv{X}[m]}]$.  Then for $0<\eps < 1$, 
\beq
\Pr_{\rv{s}}\left\{ \| \tilde{A} - A \|\geq \eps \right\} \leq 25d_1\exp\left( -10^{-11}{\eps}^3 \frac{M}{\beta}\right).
\enq 

\end{lemma}
\begin{proof}
We will rely on Corollary~\ref{cor:tononpositive}, which provides us a
similar concentration result when $d_1 \sim d_2$ (the crucial
difference is that the guarantee on $\|A_x\|_{\infty}$ is now in terms
of $d_2$, which may be much smaller than $d_1$). We will embed the
matrix $A_x$ in a $d_1 \times d_1$ matrix $B_x$. The matrix $B_x$ will
be constructed from $A_x$ in two steps. Let $d_1 = q d_2 + r$, where
$0 \leq r \leq d_2$ and $q \geq 0$ are integers.

\begin{description}
\item[Step 1 $\to$] We stack $q$ copies of $A_x$ side by side to
obtain
{
\beq
 \label{nohit}
 \tilde{A}_{x} = \frac{1}{q}
\left[\begin{array}{c|c|c c c|c}\vspace{-4mm}
 & & & &\\
 & & & &\\
 & & & &\\
 {A}_{x}&{A}_{x}& &\cdots&&{A}_{x}\\
 & & & &\\
 & & & &
\end{array}\right].
 \enq}Note that $\tilde{A}_x$ has $d_1$ rows and $\tilde{d}_2= qd_2$
columns; in particular, $d_1 \leq 2\tilde{d}_2$ as required in
Corollary~\ref{cor:tononpositive}. However, in the present form, the
values of $\tilde{A}_x$ the $\|\tilde{A}_x\|$ and
$\|\tilde{A}_x\|_{\infty}$ are not good enough to obtain the desired
concentration from Corollary~\ref{cor:tononpositive}: in particular,
the contributions from the $q$ copies of $A_x$ add up and
$\|\tilde{A}_x\|_{\infty}$ grows too large (despite the normailization
by $q$).  In order to keep the contributions from adding up, we will
apply random shifts.

\item[Step 2 $\to$] For $i \in [q]$, let $G_i$ be the following
  $d_1\times d_1$ matrix of the form $(\gamma_{ij})_{i,j=1}^d$, where
  $\gamma_{ij}$ is a complex gaussian, that is, it has the form
  $(a_{ij} + \sqrt{-1} b_{ij})/\sqrt{2d}$, where $a_{ij}, b_{ij}$ are
  chosen independently according to the standard normal distribution
  $N(0,1)$; further the random choices made for the different $G_i$
  are independent. 
With this choice of $g=\langle G_1, G_2, \ldots, G_q\rangle$,
let 
{ \beq
 \label{hit}
A^g_x = \frac{1}{q}
\left[\begin{array}{c|c|c c c|c}\vspace{-4mm}
 & & & &\\
 & & & &\\
  & & & &\\
 G_1{A}_{x}&G_2{A}_{x}&&\cdots&&G_{q}{A}_{x}\\
 & & & &\\
 & & & &
\end{array}\right].
 \enq} 
\end{description}
This completes our {\em embedding} of $A_x$ into a larger $d_1
\times \tilde{d}_2$ matrix $A_x^g$. Note that this random embedding is
determined by the random choice of $g$ (the same $g$ is used for all
$A_x$); when the same operation is performed starting with a $d_1
\times d_2$ matrix $B$, we will refer to the resulting matrix as
$B^g$.

\paragraph{The plan:} As stated before, the 
idea is to show that the concentration result of
Lemma~\ref{nonpositive} is applicable to the matrices $A_x^g$, and
conclude from this that the claimed concentrtion holds for the
original matices $A_x$. Let
\begin{eqnarray*}
\tilde{A}^g & = & \E_{m \in [M]}[A^g_{\rv{X}[m]}];\\ A^g &
= & \mathbb{E}_{\rv{X}}[A^{g}_{\rv{X}}].
\end{eqnarray*}
We thus have three tasks ahead of us (the first two help us bound
$\mu$ and $\beta$ when applying Lemma~\ref{nonpositive} to $A^g_x$,
and the third helps us conclude a concentration result for $A_x$ from
the concentration result for $A^g_x$).
\begin{enumerate}
\item[(i)] Derive an upper bound for $\|A^g_x\|$; 
\item[(ii)] Derive an upper bound for $\|A^{g}_x\|_\infty$ ; 
\item[(iii)] Relate the events 
$\cE_1 :=  \|\tilde{A} - A \|\geq \eps$ and 
$\cE'_1 := \|\tilde{A}^g - A^g \|\geq \eps'$ 
for an appropriate $\eps'$. 
\end{enumerate}

In the following set $\ell=10$ and 
$t = 4 \ln \left(\frac{480\ell^3}{\eps}\right)$.

\paragraph{(i) The upper bound for $\|A^g_x\|$:}
Consider the event
\[
\cE_{g}:= \left\{\forall x \in \universe{X}: \|A^{g}_x\| \leq \ell\|A_x\|   \right\}
\]
\begin{claim} \label{cl:tracenormub}
\begin{eqnarray*}
\Pr\{\|G_i\|_{\infty}\leq  \ell, \mbox{ for $i=1,2,\ldots,q$} \} 
&\geq & 
1 - \exp\left(\frac{-\ell^2}{16}\right);\\
\Pr\{\cE_g\} & \geq &
1-\exp\left(\frac{-\ell^2}{16}\right) \geq \frac{99}{100}.
\end{eqnarray*}
\end{claim} 
\noindent   The second inequality follows immediately from the first: if $\|G_i\|_\infty \leq \ell$
for all $i \in [q]$, then
\[ \|A^{g}_x\| \leq \frac{1}{q} \sum_{i=1}^q \|G_i\|_\infty \|A_x\| \leq 
\ell \|A_x\|.\]
To see the first inequality, observe that
\begin{align}
\Pr\{\|G_i\|_{\infty} >  \ell, \mbox{ for some $i=1,2,\ldots,q$} \}
&\overset{a} \leq q\Pr\{\|G_1\|_{\infty} > \ell \} \nonumber
\\ &\overset{b}\leq d_1\exp\left(-d_1\frac{\ell^2}{16}\right)
\label{claim:b}\\
 \label{probnorm}
 &\overset{c}\leq \exp\left(-\frac{\ell^2}{16}\right),
\end{align}
where $a$ follows from the union bound because since $G_i$s are
identically distributed, $b$ follows from Fact~\ref{fact1} below (note we
assumed $\ell =10 \geq 6$), and $c$ follows because the right hand side of 
(\ref{claim:b}) is maximum when $d_1=1$.
(End of Claim)

\paragraph{(ii) The upper bound on $\|A^g_x\|_{\infty}$:} The random
shifts applied to the matrices will be crucial in keeping
$\|A^g_x\|_{\infty}$ under control. 
Let $A$ be a $d_1 \times d_2$. To bound $\|A^g\|_\infty$, consider 
the singular value decomposition of $A$:
\beq
\label{svdA}
A = U\Lambda V^\dagger.
\enq
Consider the matrix 
\beq
\hat{A}=
  \frac{1}{q}
\left[\begin{array}{c|c|ccc|c}\vspace{-4mm}
 & & & &\\
 & & & &\\
  & & & &\\
  H_1{\Lambda}&H_2{\Lambda}& &\cdots& &H_{q}{\Lambda}\\
 & & & &\\
 & & & &\\
\end{array}\right]\left[\begin{array}{c c c c c}
V^\dagger & 0 & \cdots & \cdots &0 \\
0 & V^\dagger & \cdots & \cdots&0\\
\vdots &0 & \ddots&\cdots&\vdots\\
\vdots &\vdots & \cdots&\ddots& \vdots\\
0 &0 & \cdots&\cdots&  V^\dagger
\end{array}\right],
 \enq where $H_j = G_jU$, for $j=1,2,\ldots,q$. 
Notice that the matrices
 $\hat{A}$ and $A^g$ are identical. The matrix on the extreme right is
unitary and the norm does not change by its action. So, we focus on
\beq A^\prime=
 \frac{1}{q} \left[\begin{array}{c|c|ccc|c}\vspace{-4mm} & & &
     &\\ & & & &\\ & & & &\\ 
H_1{\Lambda}&H_2{\Lambda}&&\cdots&&H_{\frac{d_1}{d_2}}{\Lambda}\\ &
     & & &\\ & & & &
\end{array}\right].
\enq 
Next, for every $j=1,2,\ldots,q$,
let $\cT_{j}$ be an unitary such that it exchanges the $k^{th}$ row of
$\Lambda$ with its $((j-1)d_2+k)^{th}$ row, for $k=1,2,\ldots, d_2$.
Let ${H'}_j=H_j\cT^\dagger_{j}$, for  $j=1,2,\ldots,q$, and
notice that the matrix \beq
\label{prime}
  \frac{1}{q}
\left[\begin{array}{c|c|ccc|c}\vspace{-4mm}
 & & & &\\
 & & & &\\ 
 H^\prime_1\cT_1{\Lambda}&H^\prime_2\cT_2{\Lambda}&&\cdots&&H^\prime_{q}
\cT_{q}{\Lambda}\\
 & & & &\\
 & & & &
\end{array}\right],
 \enq
is identical with ${A^\prime}$. Let,
\beq
\label{Lambda}
\bar{\Lambda}= \left[\begin{array}{c c c c c c c}
\lambda_1 & 0 & \cdots & 0  & \cdots & \cdots & 0\\
0 & \ddots & 0 & \cdots &  \cdots & \cdots & 0\\
\vdots &\cdots & \lambda_{d_2}&\cdots & \cdots  & \cdots & \vdots\\
\vdots & \cdots  & \cdots & \ddots&  \cdots& \cdots & \vdots\\
\vdots & 0 & \cdots &  \cdots &  \lambda_1 &\cdots  & \vdots \\
\vdots & \cdots & \cdots & \cdots &  \cdots & \ddots  & \vdots\\
 0 & \cdots & \cdots & 0  & \cdots & \cdots & \lambda_{d_2}
\end{array}\right],
\enq 
where $\{\lambda_1, \cdots ,\lambda_{d_2}\}$ are the singular
values of the matirx $A$. It now follows that the matrix ${A^\prime}$
is equivalent in distribution with the matrix 
\beq \bar{A} =
\frac{1}{q}\bar{H} \bar{\Lambda}, \label{eq:abar} \enq 
where the
matrix $\bar{H}\in\mathbb{C}^{d_1\times d_1}$ is a random matrix with
the same distribution as the $G_i$. We will now use this reformulation
to analyse $\|A^g_x\|_\infty$.

Let \beq
\label{infnorm}
{\bf{I}}_x :=
\begin{cases}
1 & \mbox{if } \|A^g_{x}\|_{\infty}> \frac{t}{q} \|A_x\|_\infty \\
0        & \mbox{otherwise}
\end{cases},
\enq We will show that $\rv{I}_x$ is rarely $1$ and conclude that with
high probability for most $x$ (with respect to the distribution of
$\rv{X}$), $\|A^g_x\|_\infty$ is small; furthermore, we will argue that the
$x$ for which $I_x=1$ do not contribute to the sample average of
$\tilde{A}^g$.  To state this formally, let
\begin{eqnarray}
\cE_{2} &:=& \left\{\sum_{x}P_{X}(x)\mathbf{I}_x < \frac{\eps}{480\ell} \right\};
\label{E2} \\
\cE_3&:=& \left\{\left\|\E_{m \in [M]}[A^g\mathbf{I}_{\rv{X}[m]}]\right\| <
  \frac{\eps}{240} \right\}.
\label{E3}
\end{eqnarray}
\begin{claim} \label{cl:infnormub}
\begin{eqnarray}
\label{bbbbb}
\Pr\{\rv{I}_x=1\} & \leq & \exp\left(-\frac{t^2}{16}\right);\\ 
\Pr\{\cE_2\} & \geq & 1- \left(\frac{480\ell}{\eps}\right)
\exp\left(-\frac{t^2}{16}\right) \geq \frac{99}{100};\\ 
\Pr\{\cE_3 \mid \cE_2 \cap \cE_g\} &\geq& 1 - \exp\left(-2
\left(\frac{\eps}{480\ell}\right)^2 M \right). \label{ineq:sampleAg}
\end{eqnarray}
\end{claim}
For the first inequality, we carry out the above analysis with
$A:=A^g_x$ and arrive at \eqref{bbbbb} using Fact \ref{fact1}.

The second inequality follows from the first using Markov's
inequality.  

The third inequality is just a statement about
concentration of scalar sample averages near the true average, and
follows from the definition of $\cE_2$ and $\cE_3$ by a routine application of the
scalar Chernoff bound.  (End of Claim)

\paragraph{Lower bound on $\|\tilde{A}^g - A^g\|$:} We wish to show that
the probability of the event $\cE_1 := \|\tilde{A} - A \|\geq \eps$
can be bounded in terms of the event $\cE'_1 := \|\tilde{A}^g - A^g
\|\geq \eps'$ for an appropriate $\eps'$; the operators $\tilde{A}^g$
are better suited for an application of our concentration ineqalities
and we will be able to bound $\Pr[\cE'_1]$ directly. Ideally we would like 
the following event to hold:
\beq 
\cE_4 := \left\{ \|\tilde{A}^g - A^g\| \geq \frac{1}{120}\| \tilde{A} - A \|\right\}
\label{E4}
\enq 
Because we construct $A_x^g$ from $A_x$ randomly (and not deterministically), this need not always hold.
\begin{claim} \label{cl:trnormlb}
For every operator $B\in \mathbb{C}^{d_1 \times d_2}$,
\[ \Pr\{\|B^g\| \geq \frac{1}{120} \|B\|\} \geq  0.22.\]
Further, taking $\eps' = \eps/120$ in the definion of $\cE'$, we have
\[ \Pr\left\{ \cE'_1 \mid \cE_1\right\} \geq 
\Pr\{\cE_4 \mid \cE_1 \}  \geq  0.22.\]
\end{claim}
We show the first inequality in Lemma \ref{A*property2} below. The
second follows from the first because
\[  \tilde{A}^g - A^g =   (\tilde{A} - A)^g. \]
Note that $g$ is chosen independently of the random sample on
which $\tilde{A}^g$ depends.
(End of Claim)

We can now complete the proof of our lemma. Consider the event 
\[ \cE_* = \cE_1 \cap \cE_2 \cap \cE_4 \cap \cE_g.\]
Note that $\cE_2$ and $\cE_g$ depend only the random choices 
$G_i$ (and not the
sample $\rv{s}$), and are thus independent of $\cE_1$. Thus,
$E_1$ is independent of the rest. Thus, from the claims above, using
the union bound, we conclude that 
\[ \Pr\{\cE_2 \cap \cE_4 \cap \cE_g\mid \cE_1\} 
   \geq 0.22 - \frac{1}{100} - \frac{1}{100} = \frac{1}{5}. \]
Thus,          
\beq
\Pr\{\cE_*\}  \geq  \frac{1}{5}\Pr\{\cE_1\} 
\enq
Now, if $\cE_*$ holds then we have (from $\cE_1 \cap \cE_4$) that
\begin{align*}
\frac{\eps}{120}  &\leq  \left\| \E_{m \in
  [M]}[A^g_{\rv{X}[m]}]-\mathbb{E}_{\rv{X}}[A^{g}_{\rv{X}}]\right\| \\
 & \leq    
\left\| \E_{m \in
  [M]}[A^g_{\rv{X}[m]}\mathbf{I}^c_{\rv{X[m]}}]-\mathbb{E}_{\rv{X}}[A^{g}_{\rv{X}}\mathbf{I}^c_{\rv{X}}]\right\| 
 + 
\left\|
\E_{m \in
  [M]}[A^g_{\rv{X}[m]}\mathbf{I}_{\rv{X}[m]}]\right\|+
\left\|\mathbb{E}_{\rv{X}}[A^{g}_{\rv{X}}\mathbf{I}_{\rv{X}}]\right\|.
\end{align*}
Thus, one of the three terms on the right must be large; in
particular, if $\cE_*$ holds, then one of the following three events must
hold:
\begin{eqnarray*}
\cE_a &=& \left\{
\left\| \E_{m \in
  [M]}[A^g_{\rv{X}[m]} \mathbf{I}^c_{\rv{X}[m]}]-\mathbb{E}_{\rv{X}}[A^{g}_{\rv{X}}\mathbf{I}^c_{\rv{X}}]\right\|
 \geq  \frac{\eps}{480} \right\}\\
\cE_b &=& \left\{
\left\|
\E_{m \in
  [M]}[A^g_{\rv{X}[m]}\mathbf{I}_{\rv{X}[m]}]\right\|  \geq 
\frac{\eps}{240} \right\} \\
\cE_c &=& \left\{
\left\|\mathbb{E}_{\rv{X}}[A^{g}_{\rv{X}}\mathbf{I}_{\rv{X}}] \right\|\geq \frac{\eps}{480}
\right\}.
\end{eqnarray*}
We will bound $\cE_*$ by bounding each of $\Pr\{\cE_a \cap \cE_*\}$, $\Pr\{\cE_b \cap \cE_*\}$ and  $\Pr\{\cE_c \cap \cE_*\}$. Since $\cE_*$ includes $\cE_2$ and $\cE_g$, we conclude that $\cE_c
\cap \cE_*$ is impossible. Also, from Claim~\ref{cl:infnormub} (see inequality (\ref{ineq:sampleAg}))
\[ \Pr\{\cE_b \cap \cE_*\} \leq
\Pr\{\cE_b \mid \cE_2 \cap \cE_g\} \leq 
\exp\left(-2 \left(\frac{\eps}{480\ell}\right)^2 M \right).\]
We now bound $\Pr\{\cE_a \cap \cE_*\}$. Towards this notice that $\|A^g_x\mathbf{I}^c_x\|_{\infty} \leq \frac{t\beta}{d_1}$, furthermore under the assumption of the event $\cE_*$ (recall that event $\cE_*$ includes the event $\cE_g$) we have
$\|A^g_x\mathbf{I}^c_x\| \leq \ell$. We now get the desired bond by invoking Corollary~\ref{cor:tononpositive} with $d \leftarrow d_1$, $\eps \leftarrow \frac{\eps}{480}$, $A_x \leftarrow A^g_x\mathbf{I}^c_x$, $\mu \leftarrow \ell$ and $\beta \leftarrow t\beta$ and conclude that
\[ \Pr\{\cE_a \cap \cE_*\} \leq \Pr\{\cE_a\mid \cE_*\} \leq 
4d_1\exp\left(\frac{-(\eps/480)^2}{32\ln(2)\ell}\frac{M}{{2t\beta}+\ell}\right) \leq 4d_1\exp\left(\frac{-(\eps/ 480)^2}{32\ln(2)4t\ell}\frac{M}{{\beta}}\right).\]
Since $\ell = 10$ and $t = 4 \ln \left(\frac{480\ell^3}{\eps}\right)$, we conclude, that 
\beq
\frac{1}{5} \Pr\{\cE_1\} \leq \Pr\{\cE^*\} 
 \leq  
4d_1\exp\left({-10^{-11}\eps^3}\frac{M}{\beta}\right)\\
 +
\exp\left( - 10^{-8}{\eps}^2 M \right).
\enq
Our claim follows from this.
\end{proof}

\begin{lemma}
\label{A*property2}
Let $A \in \mathbb{C}^{d_1 \times d_2}$.  
Then, 
\beq
\label{desired}
  \Pr\left\{\left\|A^g\right\|\geq \frac{{1}}{120} \|A\|\right\} \geq 0.22.
  \enq
\end{lemma}
\begin{proof}
We proceed as we did above to bound $\|A^g_x\|_{\infty}$. Recall the
formulation in~(\ref{eq:abar}). Thus, it suffices to show 
\beq
\label{desired}
\Pr\left\{\left\|\bar{H} \bar{\Lambda}\right\|\geq \frac{q}{120}\|A\|\right\} \geq 0.22,
\enq
We will show the following.
\begin{claim}
\label{C11}
\beq
\Pr\left\{\left\|\bar{H} \bar{\Lambda}\right\|\geq\frac{1}{6}\tr[\bar{H}^\dagger \bar{H} \bar{\Lambda}] \right\}\geq 0.98.
\enq
\end{claim}
\begin{claim}
\label{c22}
\beq
\Pr\left\{\tr[\bar{H}^\dagger \bar{H} \bar{\Lambda}] \geq \frac{1}{20}\frac{d_1}{d_2}\left\|A\right\|\right\} \geq 0.24.
\enq
\end{claim}
Note that \eqref{desired} follows from these claims because, with
probability at least 0.22, we have
\[ \left\|\bar{H} \bar{\Lambda}\right\|\geq\frac{1}{6}\tr[\bar{H}^\dagger \bar{H} \bar{\Lambda}] \geq 
\frac{1}{120}\frac{d_1\
}{d_2}\left\|A\right\|.
\]
It remains to prove Claim \ref{C11} and Claim \ref{c22}.

{\bf{Proof of Claim \ref{C11}:}} Notice the following set of inequalities
\begin{align}
\tr[\bar{H}^\dagger \bar{H} \bar{\Lambda}] &\leq \|\bar{H}\|_{\infty}\left\|\bar{H} \bar{\Lambda}\right\|\nonumber\\
\label{probn11}
&\leq 6\left\|\bar{H} \bar{\Lambda}\right\|,
\end{align}
where \eqref{probn11} follows under the assumption of the event
$\|\bar{H}\|_{\infty}\leq 6$ which happens with probability at least
$0.98$ (see Fact \ref{fact1} below).

{\bf{Proof of Claim \ref{c22}}:} Let $\{\kappa_1,\cdots, \kappa_{d_1}\}$ represent the diagonal entries of the matrix $\bar{H}^\dagger \bar{H}$. Furthermore, from the definition of $\bar{H}$ it follows that each of the $\kappa_i$s can be represented as $\frac{\chi_i}{2d_1}$, where for every $i\in [1:d_1]$, $\chi_i$ is chi-squared distributed with $2d_1$ degrees of freedom. It is easy to see that $\tr[\bar{H}^\dagger \bar{H} \bar{\Lambda}] = \sum_{i=1}^{d_1}\frac{\chi_i\lambda_i}{2d_1}$. For every $i \in [1:d_1]$, let
\beq
\label{traceblowup1}
 \mathbf{I}_i:=
\begin{cases}
1 & \mbox{if } \chi_i \geq d_1 \\
0        & \mbox{otherwise}.
\end{cases}
\enq
Thus, it now follows from \eqref{traceblowup1} that
\beq
\label{gammaprime11}
\tr[\bar{H}^\dagger \bar{H} \bar{\Lambda}] \geq \frac{1}{2}\sum_{i=1}^{d_1}\mathbf{I}_i\lambda_i.
\enq
Notice that
\beq
\label{expectmarkov1}
\mathbb{E}\left[{\sum_{i=1}^{d_1}\mathbf{I}_i\lambda_i}\right]
 = \Pr\left\{\chi_i \geq d_1\right\} {\|{\bar{\Lambda}}\|}.
\enq
By Fact 1 (b) below (with $\beta = \frac{1}{2}$ and $d_1 \geq 2$), we obtain
\beq
\label{expect1}
\Pr\left\{\chi_i \geq d_1\right\} > 0.32,
\enq
implying that 
\[ \mathbb{E}\left[\sum_{i=1}^{d_1}\mathbf{I}_i\lambda_i \right] \geq
   0.32 {\|{\bar{\Lambda}}\|}.\]
Thus, since ${\sum_{i=1}^{d_1}\mathbf{I}_i\lambda_i} \leq \|{\bar{\Lambda}}\|$,
we have from Fact 1 (c) that
\begin{align}
\label{lowerb11}
\Pr\left\{{\sum_{i=1}^{d_1}\mathbf{I}_i\lambda_i} > 0.1 {\|\bar{\Lambda}\|}\right\} > 0.24.
\end{align}
The claim now follows from \eqref{gammaprime11} and \eqref{lowerb11}. This completes the proof.
\end{proof}

\begin{fact}
\label{fact1}
\begin{enumerate}
\item[(a)]
For $i,j=1,2,\ldots, d$, let the random complex number $\gamma_{ij} =
(a_{ij} + \sqrt{-1} b_{ij})/\sqrt{2d}$ be such that $a_{ij}, b_{ij}$
are chosen independently according to the normal standard distribution
$N(0,1)$.  Let $G\in \mathbb{C}^{d_1\times d_1}$ be the random matrix
$(\gamma_{ij})_{i,j=1}^d$. Then, for $\ell \geq 6$,
 \beq \Pr\{\|G\|_{\infty} \geq \ell\} \leq
 \exp\left(-\frac{d \ell^2}{16}\right). 
  \enq
\item[(b)] Let $X= \eta_1^2 + \eta_2^2 + \cdots + \eta_{2d}^2$ where
  each $\eta_i$ is chosen independently with distribution $N(0,1)$
  (that is, $X$ has chi-squared distribution with $2d$ degrees of
  freedom). Then, for $0<\beta<1$ \beq \Pr\{X<2\beta d\} \leq (\beta
  e^{1-\beta})^{{d}}. \nonumber \enq
  \item [(c)] Let $X$ be a positive random variable that satisfies $\Pr\{X\leq \alpha\}= 1$ for some constant $\alpha$. Then for $c \leq \mathbb{E}[X]$,
\beq
\Pr\{X>c\} \geq \frac{\mathbb{E}[X]-c}{\alpha-c} \nonumber.
\enq

\end{enumerate}
\end{fact}

\begin{proof}
\item[(a)] See \cite[Fact 6]{Sen-2006}
\item[(b)] The moment generating function of $X$ is given by
  $\E[e^{tX}] = (1-2t)^{-d}$ for $t < \frac{1}{2}$ \cite{ross-2009}.  Set $t= - \frac{1-\beta}{2\beta} < 0$ and observe that
$e^{tx} > e^{2t\beta d}$, whenever $x \leq 2\beta d$. 
By Markov's inequality
\beq \Pr\{X<2\beta d\}  \leq  (1-2t)^{-d}e^{-2t\beta d} =
(\beta e^{1-\beta})^{{d}}. \nonumber
\enq
\item [(c)] This claim follows from Markov's inequality. To see this let $\hat{X} = \alpha-X$. It now follows that $\Pr\left\{X \leq c\right\} = \Pr\left\{\hat{X} \geq \alpha-c\right\} \leq \frac{\mathbb{E}\left[\hat{X}\right]}{\alpha-c} =\frac{\alpha-\mathbb{E}\left[{X}\right]}{\alpha-c}$. 
\end{proof}

\bibliographystyle{ieeetr}
\bibliography{master}

\begin{thebibliography}{10}

\bibitem{arbitrary-wiretap}
{M. Bloch} and {J. N. Laneman}, ``On the secrecy capacity of arbitrary wiretap
  channel,'' in {\em Proc. Allerton Conf. Commun. Control, Computing},
  (Monticello, IL, USA), Sept. 2008.

\bibitem{han-book}
{T. S. Han}, {\em Information-Spectrum Methods in Information Theory}.
\newblock Berlin, Germany: Springer-Verlag, 2003.

\bibitem{devetak-2005}
{I. Devetak}, ``The private classical capacity and quantum capacity of a
  quantum channel,'' {\em IEEE Trans. Inf. Theory}, vol.~51, pp.~44--55, Jan.
  2005.

\bibitem{cai-winter-yeung}
{N. Cai}, {A. Winter}, and {R. W. Yeung}, ``Quantum privacy and quantum
  channel,'' {\em Prob. Inf. Trans.}, vol.~40, no.~4, pp.~318--336, 2004.

\bibitem{wang-renner-prl}
{L. Wang} and {R. Renner}, ``One-shot classical-quantum capacity and hypothesis
  testing,'' {\em Phys. Rev. Lett.}, vol.~108, pp.~200501--200505, May 2012.

\bibitem{wilde-book}
{M. M. Wilde}, ``From classical to quantum {Shannon} theory.''
  http://arxiv.org/abs/1106.1445, 2011.

\bibitem{ahlswede-02-converse}
{R. Ahlswede} and {A. Winter}, ``Strong converse for identification via quantum
  channels,'' {\em IEEE Trans. Inf. Theory}, vol.~48, pp.~569--579, Mar. 2002.

\bibitem{Hayashi-noniid}
{M. Hayashi} and {H. Nagaoka}, ``General formulas for capacity of
  claasical-quantum channels,'' {\em IEEE Trans. Inf. Theory}, vol.~49,
  pp.~1753--1768, 2003.

\bibitem{datta-2014-secondorder}
{N. Datta} and {F. Leditzky}, ``Second-order asymptotics for source coding,
  dense coding and pure-state entanglement conversions.'' arXiv:1403.2543v3,
  June 2014.

\bibitem{renes-renner-2011}
{J. M. Renes} and {R. Renner}, ``Noisy channel coding via privacy amplification
  and information reconciliation,'' {\em IEEE Trans. Inf. Theory}, vol.~57,
  pp.~7377--7385, Nov. 2011.

\bibitem{Pinsker-book}
{M. S. Pinsker}, {\em Information and Information Stability of Random Variables
  and Processes}.
\newblock Holden Day, 1964.

\bibitem{tropp-2011}
{J A. Tropp}, ``User-friendly tail bounds for sums of random matrices.''
  arXiv:1004.4389, June 2011.

\bibitem{winter-99-converse}
{A. Winter}, ``Coding theorem and strong converse for quantum channels,'' {\em
  IEEE Trans. Inf. Theory}, vol.~45, pp.~2481--2485, Nov. 1999.

\bibitem{bhatia-1997}
{R. Bhatia}, {\em Matrix Analysis}.
\newblock New York: Springer-Verlag, 1997.

\bibitem{Sen-2006}
{P. Sen}, ``Random measurement bases, quantum state distinction and
  applications to the hidden subgroup problem,'' in {\em Proc. IEEE Conf. on
  Comput. Complexity}, (Prague), July 2006.

\bibitem{ross-2009}
{S. Ross}, {\em A First Course in Probability}.
\newblock Prentice Hall, 2009.

\end{thebibliography}

\end{document}